\documentclass[]{interactPublished}

\usepackage[UKenglish]{babel}

\usepackage{epstopdf}
\usepackage{subfigure}

\usepackage[numbers,sort&compress,merge]{natbib}
\bibpunct[, ]{(}{)}{,}{n}{,}{,}
\renewcommand\bibfont{\fontsize{10}{12}\selectfont}
\renewcommand\citenumfont[1]{\textit{#1}}
\renewcommand\bibnumfmt[1]{(#1)}

\theoremstyle{plain}

\theoremstyle{definition}

\theoremstyle{remark}

\usepackage{amsmath}
\usepackage{graphicx}
\usepackage[colorinlistoftodos]{todonotes}
\usepackage[colorlinks=true, allcolors=blue]{hyperref}
\usepackage{tabularx}

\newcommand{\IP}{I_{\mathrm{p}}}
\newcommand{\etal}{\textit{et al.}\,}

\usepackage{enumerate}
\usepackage{wasysym}

\begin{document}

\articletype{Review}

\title{Attoclock revisited on electron tunnelling time}

\author{
	\name{C.~Hofmann\textsuperscript{a,b}\thanks{CONTACT C. Hofmann. Email: chofmann@pks.mpg.de}, A.~S.~Landsman\textsuperscript{b,c} and U.~Keller\textsuperscript{a}}
	\affil{\textsuperscript{a}Physics Department, ETH Zurich, 8093 Zurich, Switzerland; \textsuperscript{b}Max Planck Institute for the Physics of Complex Systems,  N\"{o}thnitzer Stra\ss e 38, D-01187 Dresden, Germany; \textsuperscript{c}Department of Physics, Max Planck Postech, Pohang, Gyeongbuk 37673, Republic of Korea}
}

\maketitle

\begin{abstract}
The last decade has seen an intense renewed debate on tunnelling time, both from a theoretical and an experimental perspective.
Here, we review recent developments and new insights in the field of strong-field tunnel ionization related to tunnelling time, and apply these findings to the interpretation of the attoclock experiment  [Landsman et al., \emph{Optica 1}, 343 (2014)].
We conclude that models including finite tunnelling time are consistent with recent experimental measurements. 
\end{abstract}

\begin{abbreviations}
	
	\begin{tabularx}{0.9\textwidth}{lX}
		A &  adiabatic \\
		ADK &  Ammosov, Delone and Krainov model \cite{ADK1986,Delone1991} \\
		CEO &  Carrier-Envelope-Offset phase $\varphi_{\mathrm{CEO}}$ \\
		CoM	&  centre of mass \\
		CTMC &  Classical Trajectory Monte Carlo simulation \\
		FWHM &  full width half maximum \\
		IR &  infrared  \\
		KR &  Keldysh-Rutherford model \\
		NA &  non-adiabatic \\
		PMD &  Photoelectron Momentum Distribution \\
		PPT &  Perelomov, Popov and Terent'ev model \cite{PPT1,PPT2} \\
		SAE &  Single Active Electron approximation \\
		SCT &  Single Classical Trajectory \\
		SFA &  Strong Field Approximation \\
		TDSE &  Time-Dependent Schr\"{o}dinger Equation \\
	\end{tabularx}
\end{abbreviations}


\tableofcontents

\clearpage
\section{Introduction} \label{sec:Intro}

Quantum tunnelling is a fundamental and ubiquitous process that sparked a long-standing debate on its duration \cite{Landauer1989,Landsman2015} since the concept was first conceived \cite{MacColl1932}.
Time is not an operator in quantum mechanics, but rather a parameter in the time-dependent Schr\"{o}dinger equation (see for example \cite{Pauli1980} p. 63). 
This fact is often used as a throw-away argument claiming that in consequence, the question ``how long does it take for a quantum particle to tunnel through a potential barrier'' is not physically valid.
On the other end of the debate scale, there is the notion that it should be ``easy, just follow the peak of the wave packet''. 
The peak of the wave packet is the relevant observable when determining the group delay of a dispersive wave packet
\begin{equation}
T_g = \frac{z}{v_g} = z\cdot  \frac{dk}{d\omega} = \frac{d\phi}{d\omega},
\end{equation}
where $v_g$ is the group velocity, $\phi$ is the phase of the wave packet for a particular energy component $\omega$, and $k$ is the corresponding wave number.

\begin{figure}[hb]
	\centering
	\includegraphics[width=0.5\linewidth]{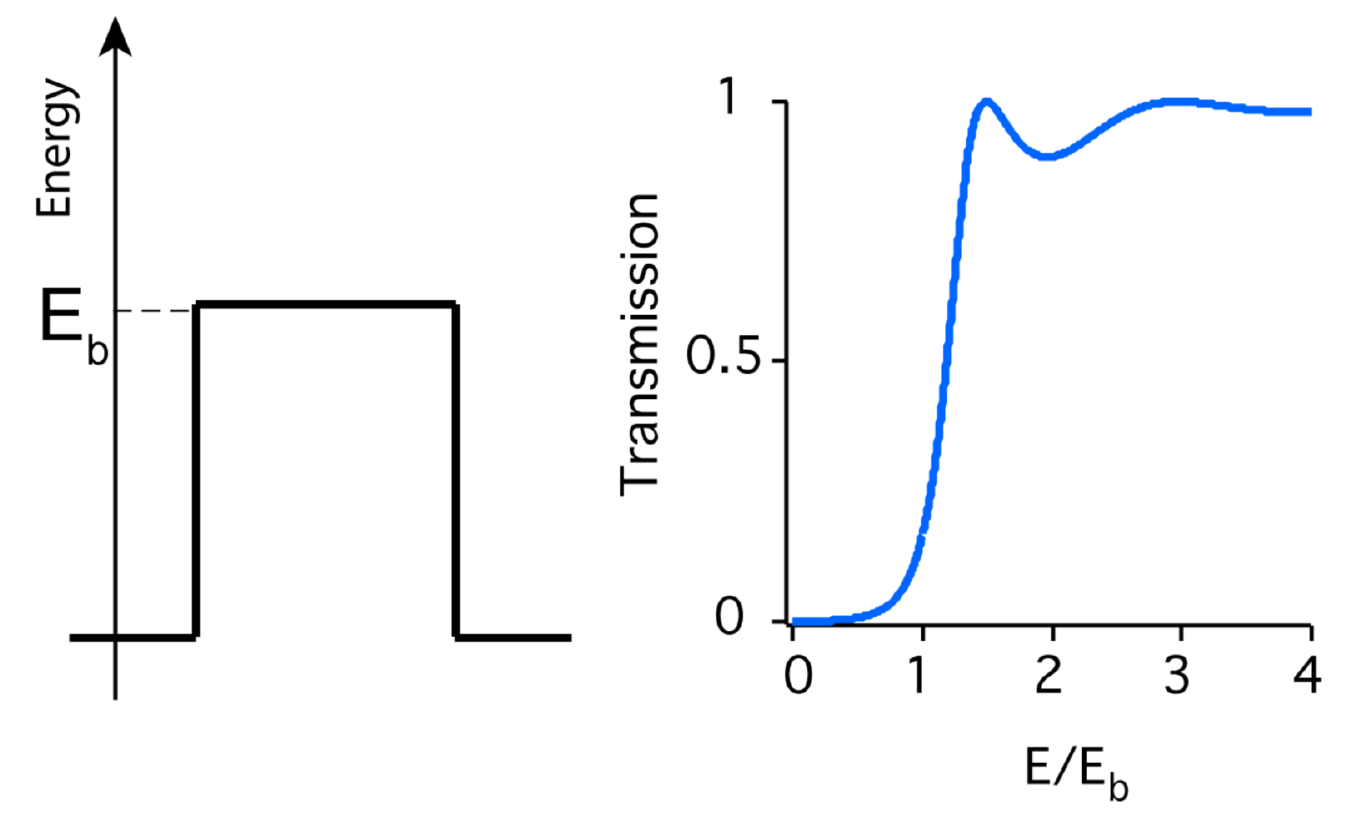}
	\caption{A potential barrier acts as a high-pass filter for the wave packet, thus strongly modifying the energy components of the ionised wave packet.}
	\label{fig:TransmissionFilter}
\end{figure}

The  Wigner delay $\tau_W$, often applied to ionization delays, \cite{Isinger2017} (see also section \ref{sec:SPI}) formally corresponds to the group delay,
\begin{equation}
\tau_W = \hbar \frac{d\phi}{dE} = \frac{d\phi}{d\omega} = T_g. \label{eq:WignerDelayGroupDelay}
\end{equation}
However, this concept depends on the fact that the spectrum of the wave packet is unchanged -- a condition not satisfied in the tunnelling process. 
In particular, tunnelling acts as an energy filter, favouring higher-energy components of the incident wave packet, see figure \ref{fig:TransmissionFilter}. 
Or in the words of M. Buettiker: "There is no conservation law for the peak of a wave packet."

Additionally, the electron wave packet is chirped during the propagation in vacuum, unlike photon wave packets.
The combination of this chirp with the energy filtering during the tunnelling process means that the Wigner formalism \cite{Isinger2017} for ionization delays is not applicable to the tunnelling ionization case \cite{Sabbar2015,Sabbar2017erratum,Gallmann2017}, where a valence electron tunnels through a potential barrier created by the superposition of the binding Coulomb potential with a strong laser field.

The attoclock is a recently developed approach for the extraction of tunnelling time in the context of strong field ionization \cite{Eckle2008,Eckle2008a}.
The most recent attoclock experimental measurements \cite{Landsman2014b}, which found sub-luminal tunnelling times over a wide intensity range, sparked a number of theoretical developments 
\cite{Zimmermann2016,Ni2016,Teeny2016,Torlina2015}.
Two other independent attoclock experiments \cite{Camus2017,Sainadh2017} recently came to opposite conclusions regarding the duration of the tunnelling process. Additionally, an experiment on rubidium atoms tunnelling in a kicked optical lattice \cite{Fortun2016} also found finite tunnelling times on a much slower timescale of microseconds, due to the much heavier particles involved. 
It seems that ultrafast laser technology finally enabled experiments to provide evidence supporting a quote by Landauer in 1989 \cite{Landauer1989}: 
\begin{quote}
	More important than the exact result and its relation to theoretical controversies, is the fact that a timescale associated with the barrier traversal can be measured, and is a real (not imaginary) quantity.
\end{quote}
While most experiments seem to agree that quantum tunnelling does not happen instantaneously, there is no consensus yet on the most recent theoretical side \cite{Zimmermann2016,Torlina2015,Teeny2016,Sainadh2017,Yuan2017a,Wang2018a,Ni2018a,Bray2018b,Ivanov2018}.

Here, we discuss the implications of recent new discoveries on the interpretation of attoclock experiments, as well as compare the variety of approaches used to extract tunnelling times in strong field ionization. 
This topic is important not only to the interpretation of time-resolved studies in attosecond physics, but also in the treatment of many experimental schemes in the atomic and molecular optical physics community which are based on a semiclassical view of strong-field ionization \cite{Meckel2008,Lin2010,Bruner2015}. 

\label{sec:Terminology}
For the sake of clarity, we will use the following terminology. \\
\begin{description}
	\item[transition point $t_s$:] The transition point $t_s$ is a complex moment in time, usually determined in a Strong Field Approximation (SFA) calculation as the saddle point time, and sometimes interpreted as the beginning of the tunnelling process \cite{Dykhne1960,Dykhne1962,PPT1,PPT2,Yudin2001,Anatomy,Klaiber2015}. 
	\item[starting time $t_0$:] The starting time $t_0$ conceptually corresponds to the real part of the transition point, $\Re[ t_s]$, meaning the beginning of the quantum tunnelling process on the real time axis. 
	\item[ionization time $t_i$:] The ionization time $t_i$ denotes the moment in time when an electron wave packet appears in the continuum.
	It is typically real-valued \cite{PPT2,Klaiber2015}.
	\item[tunnelling time $\tau$:] The tunnelling time $\tau = t_i - t_0$ describes the potential barrier traversal time, or in other words, the duration of the tunnelling process. 
	\item[attoclock delay $\tau_A$:] The attoclock delay $\tau_A$ describes the tunnelling time as defined in the attoclock method, $\tau_A = t_i - t_0$, where $t_0$ is assumed to be the moment when the electric field is maximized \cite{Eckle2008,Eckle2008a,Pfeiffer2012}, and $t_i$ is reconstructed from the measurements \cite{Pfeiffer2012,Landsman2014b}.

\end{description}

\subsection{Attoclock Experiment} \label{sec:AttoclockExperiment}

The strong-field ionization process encodes the moment when an electron is entering the continuum in the final asymptotic kinetic momentum $\mathbf{p}$ of the photoelectron measured on a detector \cite{Krausz2009}.
This is due to the conservation of the canonical momentum
\begin{equation}
\mathbf{p} = \mathbf{v}(t)+e\mathbf{A}(t), \label{eq:ConservationCanonicalMomentum}
\end{equation} 
where $\mathbf{v}$ denotes the velocity of a photoelectron at time $t$, and $\mathbf{A}$ the vector potential at the same time. 
This conservation law is valid under the assumption that during the propagation of the freed electron, the influence of the parent ion Coulomb force can be neglected (Strong Field Approximation SFA).
Throughout the paper, atomic units (au) are used unless otherwise specified. 

\begin{figure}[htb]
\centering
\includegraphics[width=0.7\linewidth]{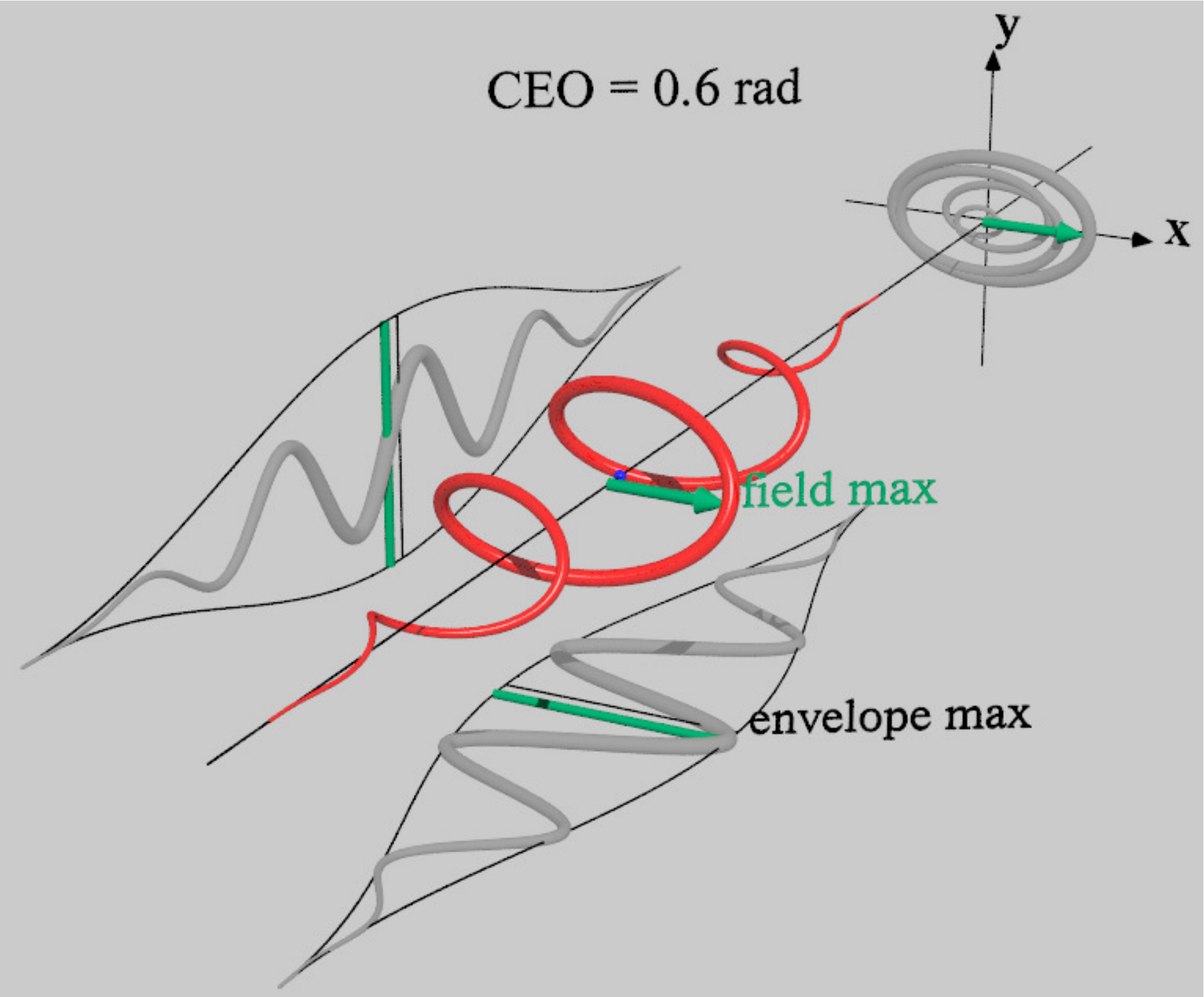}
\caption{Example for a pulse wave form in the attoclock experiment. The field is elliptically polarized with $x$ as the major polarization axis and $y$ the minor axis. The envelope reaches its maximum value for $t=0$, but the field maximum might be shifted due to the carrier-envelope-offset (CEO) phase $\phi_{\mathrm{CEO}}$.}
\label{fig:PovRayInputTemplate}
\end{figure}

At the core of the attoclock experiment \cite{Eckle2008,Eckle2008a,Pfeiffer2012,Cirelli2013,Landsman2014b} lies the comparison of experimentally observed final momenta with calculated values from a semiclassical strong-field tunnel ionization model. 
For the measurement, ellipticity $\epsilon=0.87$, helium atoms as targets and a near-infrared (near-IR) wavelength of $\lambda = 735\,\mathrm{nm}$ were chosen \cite{Eckle2008a,Landsman2014b}. This results in a rotating electric field with a rotation period of approximately $2.7\,\mathrm{fs}$, see figure \ref{fig:PovRayInputTemplate} for an example sketch.
The wave form used in the attoclock experiment can be described as
\begin{equation}
\mathbf{F}(t) = \frac{F_0}{\sqrt{1+\epsilon^2}}\left(  \cos(\omega t + \phi_{\mathrm{CEO}})\hat{x} - \epsilon\sin(\omega t + \phi_{\mathrm{CEO}})\hat{y}  \right)\cdot f(t), \label{eq:LaserField}
\end{equation}
where $F_0 = \sqrt{I}$ is the field strength constant related to the peak intensity $I$, $\omega = 0.062\, \mathrm{au}$ the angular frequency related to the central wavelength $\lambda = 735\,\mathrm{nm}$, major axis of polarization along $x$, and propagation along $z$ direction. The pulse envelope $f(t)$ with $f(0)=1$ defines a pulse duration of 6 fs (7 fs) FWHM for the lower (higher) intensity regime respectively. For our simulations (see section \ref{sec:ClassicalTrajectories}) we used a $\cos^2$ shaped envelope.
The carrier-envelope-offset (CEO) phase $\phi_{\mathrm{CEO}}$ was not stabilized in the experiment \cite{Telle1999}, to prevent any artificial angular shifts due to stabilization fluctuations \cite{Eckle2008a,Smolarski2010a}. 
This leads to random $\phi_{\mathrm{CEO}}$ for each pulse. 
The maximal field amplitude is therefore
\begin{equation}
F_{\mathrm{max}} = \frac{F_0}{\sqrt{1+\epsilon^2}}\cdot f(|\phi_{\mathrm{CEO}}/\omega|).
\end{equation}
It was shown in \cite{Eckle2008a,Eckle2008,Smolarski2010a} that a randomized CEO phase averages out to an effective $\phi_{\mathrm{CEO}} = 0$ for the observable of the most probable final momentum. 
This is due to the strong dependence on the absolute field strength of the ionization probability.
Since the CEO phase was not stabilized in the experiment, corresponding calculation must either integrate over a random distribution of CEO phases as well, or be executed for the averaged effect of $\phi_{CEO} = 0$ \cite{Eckle2008a}. 
The attoclock analysis of the experiment is only concerned with the most probable final momentum, or the highest probability density value \cite{Eckle2008a,Landsman2014b}.
From now on, we assume $\phi_{CEO} = 0$ in all calculations.

\begin{figure}[ht]
\centering
\includegraphics[height=0.38\linewidth]{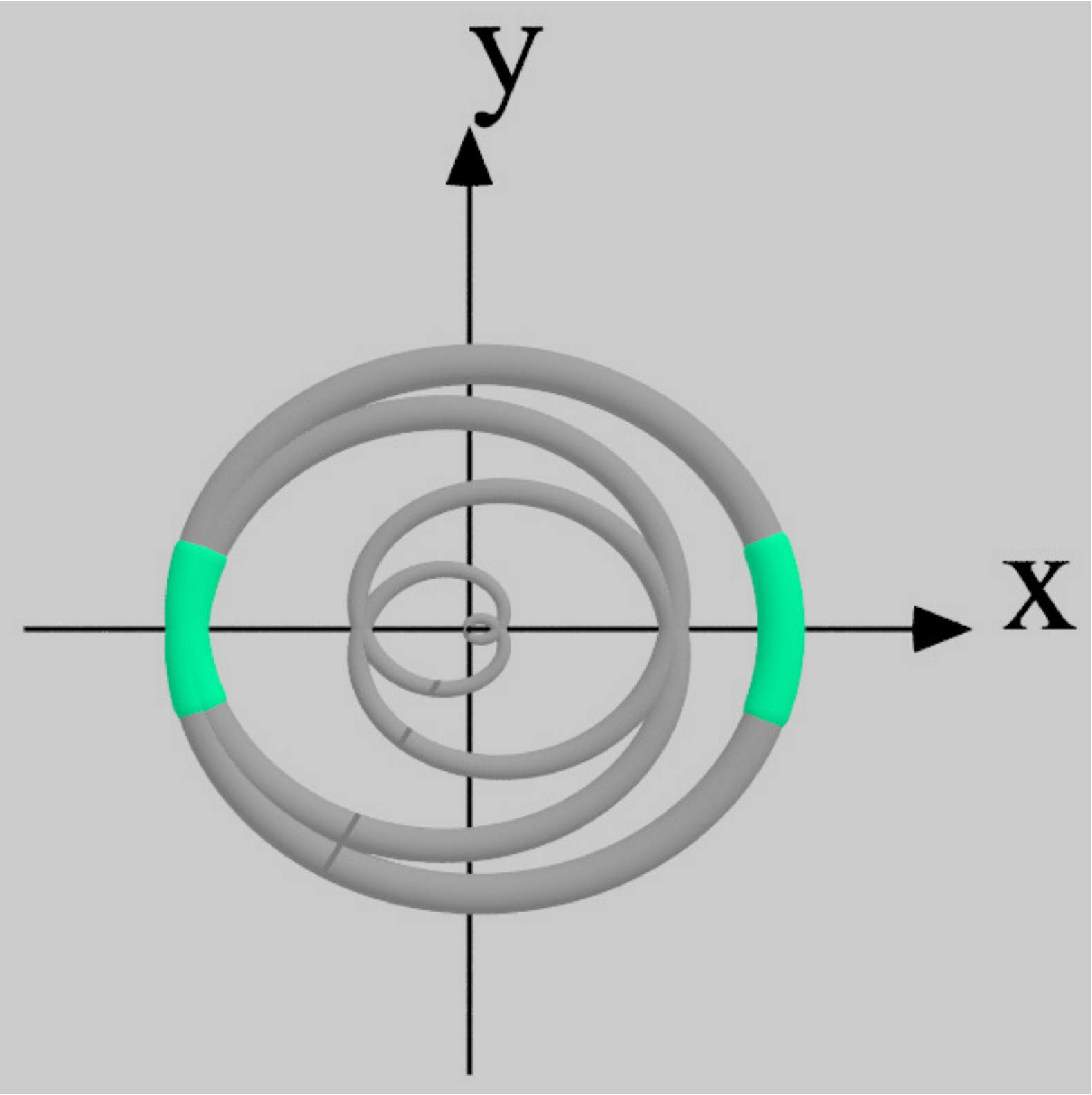} \hfill 
\includegraphics[height=0.4\linewidth]{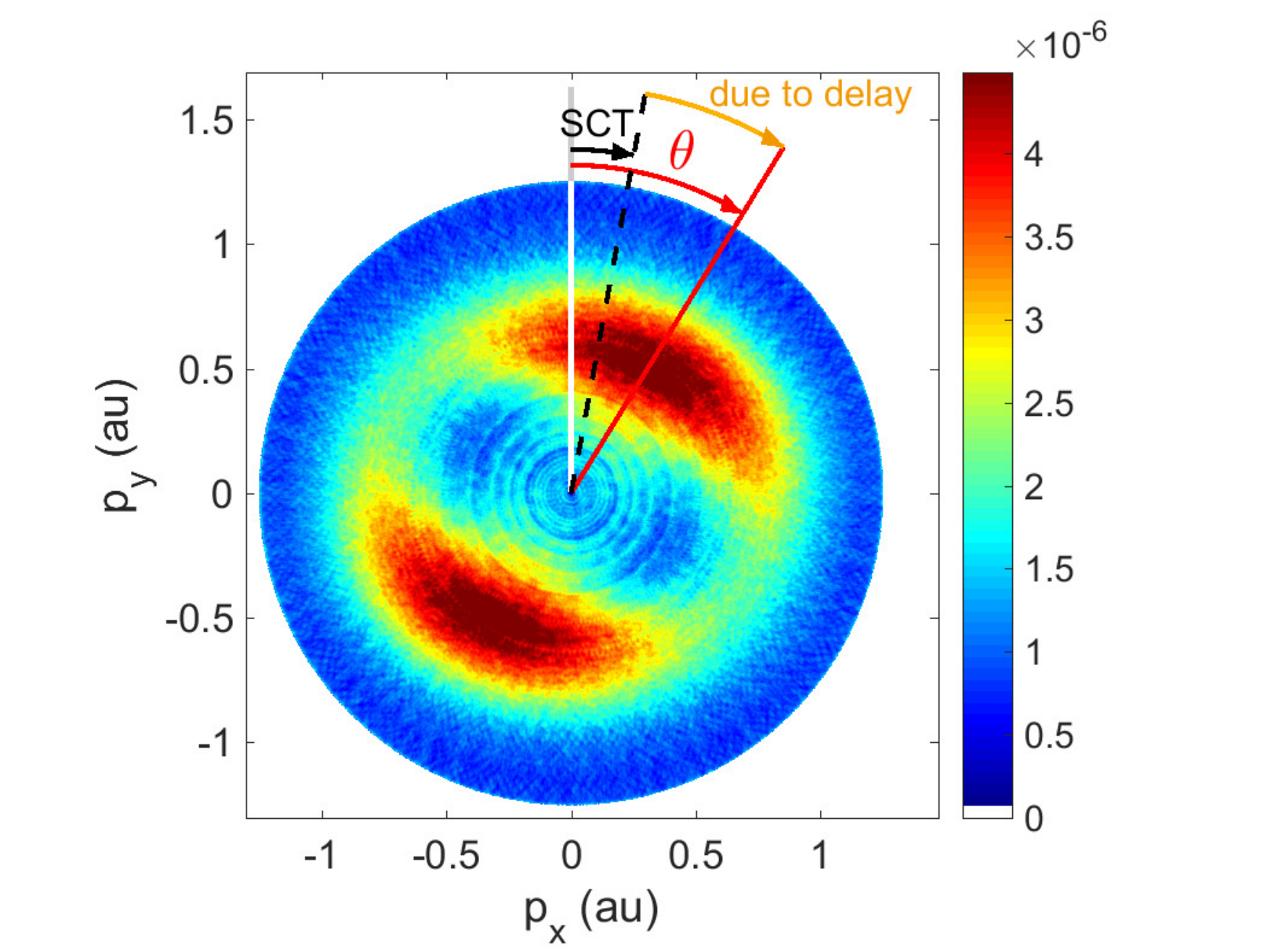} 
\caption{\textbf{Photoelectron momentum distribution (PMD):} Example of a PMD in the attoclock experiment, projected onto the polarization plane $xy$ \cite{Landsman2014b}. 
The major axis of polarization is along the $p_x$-axis. 
According to \eqref{eq:ConservationCanonicalMomentum}, the majority of photoelectrons should therefore have final momentum along the $p_y$-axis.
The red line marks the the final electron momentum direction with the highest photoelectron count rate, which corresponds to the most probable photoelectron trajectory. Any streaking angle deviating from 90 degrees (marked by the perpendicular white/gray line) consitutes an offset angle $\theta$, measured in the rotation direction of the laser field. Here, the measured offset angle $\theta$ is larger than the predicted streaking angle assuming instantaneous tunnelling (marked as a black dashed line) from a single classical trajectory (SCT) calculation.
}
\label{fig:VMIdataSCT}
\end{figure}

The aforementioned conservation of canonical momentum is exploited by comparing the measured final momentum offset angle in the plane of polarization $\theta$ (see figure \ref{fig:VMIdataSCT}) to calculations assuming that the free propagation starts (for the most probable electron trajectory) exactly at the peak of the electric field $t_0 = 0$ \cite{Eckle2008a,Landsman2014b}. 
This zero-time assumption of $t_0 = 0$ means that a polarization measurement determines the orientation of the polarization ellipse in the laboratory frame, yielding the reference for the streaking angle measurement, compare figures \ref{fig:PovRayInputTemplate} and \ref{fig:VMIdataSCT}.

Consequently, the conclusions of the attoclock experiment depend on the characteristics of the zero-time reference, and the approximations going into it.
These calculations were performed in a semiclassical framework, where an analytical calculation of the quantum tunnelled wave packet describes the probability distribution of initial conditions for classical trajectories. For a Classical Trajectory Monte Carlo (CTMC) simulation, this probability distribution is sampled for a cloud of trajectories, which then mimic the propagation of the electron wave packet after ionization \cite{Ehrenfest1927}.
Taking only the most probable initial conditions for all parameters results in a Single Classical Trajectory (SCT). 
The SCT follows the highest probability density of the ionized wave packed, see section \ref{sec:SCTvsCTMC} for a detailed discussion.
The classical trajectory numerical method allows to fully take account of the ion Coulomb force superposed with the strong laser field during the propagation \cite{Landsman2013b}, as well as other effects such as an induced dipole in the parent ion \cite{Pfeiffer2012}.

The assumptions and approximations included in the complete attoclock experiment analysis are as follows.
\begin{enumerate}[(i)]
\item \emph{Dipole approximation}: the spatial dependence of the laser field is neglected, requiring that the wavelength is much larger than the target size, and the Lorentz force induced by the magnetic field to be negligibly small \cite{Reiss2014,Ludwig2014,Danek2018,Maurer2018}. 
Also, the laser pulse is short enough that the electron does not travel any significant distance out of the focus before the pulse has already finished. 
\item \emph{Single Active Electron (SAE) approximation}: it is assumed that the helium target atoms are only singly ionized, and the second electron remains in it's (ionic) ground state. Furthermore, the approximation neglects any electron-electron interactions. Instead, it uses an effective Coulomb potential assuming that the remaining bound electron screens the ion perfectly \cite{Emmanouilidou2015,Majety2017}.
\item \emph{Adiabatic (A) approximation} or \emph{non-adiabatic (NA) framework}: In the adiabatic approximation, it is assumed that the temporal change of the laser field is relatively slow compared to the response time of the bound wave function, such that the wave function can instantaneously adapt. This also implies that the tunnelling process can be calculated in a quasistatic picture. 
On the other hand, in the non-adiabatic framework the temporal dynamics of the laser field and thus the temporal changes to the binding potential of the atom are considered. This has several consequences, including that the tunnelling electron gains some energy from the oscillating or rotating field \cite{PPT2,Mur2001,Hofmann2014,Ni2018}.
\item \emph{Classical trajectories} mimicking the propagation of a quantum wave packet: Classical dynamics agree exactly with quantum dynamics as long as the spatial dependence of the driving potential is a polynomial of second order or lower \cite{Ehrenfest1927}. 
This is the case within the SFA, but not any more if the weak influence of the Coulomb potential is accounted for. 
However, as long as these classical trajectories stay far enough away from their parent ion, the quantum correction is negligible and the classical dynamics can represent the propagation of the photoelectron wave packet \cite{Shvetsov-Shilovski2013}.
\item \emph{"Zero-time" estimate} $t_0$: in order to derive a duration of the tunnelling process, an estimate for the beginning moment is required. In the attoclock analysis, the most probable starting point for tunnelling is assumed to be when the field strength is the strongest, corresponding to the shortest tunnelling barrier.
\end{enumerate}

In the forthcoming sections, we will take a closer look at different approximations. Recent research on their validity is presented, and implications for the interpretation of strong field ionization experiments in general and the attoclock experiment in particular are discussed.

\subsection{Comparison with single photon ionization} \label{sec:SPI}
Attosecond photo ionziation delays in atoms have been first measured in the tunnel ionization \cite{Eckle2008a} 
and then in the single-photon ionization regime \cite{Schultze2010}. 
More detailed measurements and theory confirmed that in the simplest case, when the electron is promoted into a flat (non-resonant) continuum by direct single photon ionization, the corresponding ionization delay is then given by the Wigner delay, which can be expressed as the energy derivative of the scattering phase and is equivalent to the group delay of the departing electron wave packet \cite{Isinger2017}, see also \eqref{eq:WignerDelayGroupDelay}. 
To date different attosecond techniques have confirmed this result taking into account a measurement induced delay \cite{Dahlstrom2012,Nagele2011,Pazourek2015a}. 
This is in contrast to the tunnel ionization where our experimental results do not correspond to the Wigner delay because the center of wave packet makes a phase jump when a chirped wave packet propagates with an energy filter \cite{Gallmann2017,Landsman2014b,Sabbar2015,Sabbar2017erratum} 
(see Section \ref{sec:Intro}). 
In this case we loose the direct link to the classical trajectory with the centre of the electron wave packet following the Ehrenfest's theorem \cite{Ehrenfest1927}. 
However with a flat continuum we do not have such an energy filter and ionization delay is correctly described by the Wigner delay. 
The situation becomes more complicated when ionization occurs in the vicinity of autoionizing states which significantly affect the Wigner delay \cite{Sabbar2015,Sabbar2017erratum}. 
This was further confirmed most recently with angle and spectrally resolved measurements where we could demonstrate in collaboration with Anne L`Huillier that not only the phase of the photoelectron wave packet is significantly distorted in the presence of these autoionization resonances in argon, but that this distortion also depends on the electron emission angle \cite{Cirelli2018}. 
In this situation again we loose the direct link between the Wigner delay and the classical trajectory of the liberated electron.

Angular streaking was initially applied to attosecond pulse measurements \cite{Constant1997,Zhao2005} before we applied it to the attoclock concept \cite{Eckle2008,Eckle2008a}. To characterize the temporal structure of ultrafast free electron pulses \cite{Hartmann2018,Schweizer2018} the ultrafast X-ray pulse promotes electrons of a target gas into the continuum by single photon ionization, and these photoelectrons are subsequently streaked by a close to circularly polarized pulse of longer wavelength. However moving away from a pump-probe scheme with circular polarization to a single pulse with elliptical polarization was the key idea to obtain a self-referencing ``time-zero'' calibration for the attoclock \cite{Eckle2008a}. These ideas then for example also have been applied to measure the time-dependent polarization of an ultrashort pulse with sub-cycle resolution \cite{Boge2014a}.

\subsection{Other experiments on tunnelling delay}
Following the attoclock measurements performed in the Keller group \cite{Eckle2008,Eckle2008a,Pfeiffer2012,Boge2013,Landsman2014b}, a number of other experimental groups measured tunnelling time.
A completely different approach outside the ultrafast physics community was pursued by Fortun and coworkers \cite{Fortun2016}. 
They studied rubidium atoms trapped in an optical lattice tunnelling from one potential well to the next, when the lattice is suddenly kicked. 
The authors came to the conclusion that the atoms experienced a tunnelling time of the order of microseconds across potential barriers of width on the order of nanometers, since the tunnelled wave packets seemed to lag behind the reflected wave packets in their oscillation inside the neighbouring lattice cell \cite{Fortun2016}.

An experimental-theoretical collaboration published their results \cite{Camus2017} comparing the attoclock observable of final momentum direction $\theta$ between two different target species, argon and krypton. 
They too found that a quantum calculation based on the Eisenbud-Wigner-Smith approach \cite{Wigner1955}, including both a finite real tunnelling time as well as an initial longitudinal momentum, 
reproduced their measurements, whereas calculations assuming instantaneous tunnelling failed to do so even qualitatively \cite{Camus2017}. 
Classical trajectories reproducing their measurements were not only required to start at a time $t_i>0$ after the peak of the pulse, they were also required to have some positive longitudinal momentum. 
An important feature of this experiment is the fact that the conclusions do not depend on the field strength calibration (see \cite{Boge2013,Hofmann2014,Hofmann2016,Cai2017} and section \ref{sec:NA_CTMC} for more details on this issue), since the observables are directly compared with respect to the average absolute momentum. 
On the other hand, the authors assume that the SAE approximation is also valid for both argon and krypton targets, where the ionization happens out of 3p or 4p orbitals. 
Multi-electron effects in helium will be discussed in section \ref{sec:SAEtest}.

More recently, Sainadh \etal published an attoclock measurement on atomic hydrogen, comparing their experimental data to time-dependent Schr\"{o}dinger Equation (TDSE) calculations \cite{Sainadh2017}. 
They found that their codes reproduce the experimental values when the Coulomb potential is included, and yield zero streaking offset angle when a Yukawa short range potential is employed, in agreement with prior findings \cite{Torlina2015}.
This result was used by the authors as evidence of instantaneous tunnelling time in hydrogen \cite{Sainadh2017}.

\subsection{More general concepts} \label{sec:GeneralConcepts}

Apart from the above discussed approximations and calculation concepts affecting the attoclock interpretation, there are a few more which are commonly found in strong-field ionization models. 

For most analytical calculations, the binding potential of the target atom is approximated as a short-range potential. This can mean that the extreme case of a delta-potential is used \cite{PPT2}, or a Yukawa potential exponentially suppressing the long range Coulomb tale \cite{Torlina2015,Sainadh2017,Pazourek2015a}. 
For the propagation of a freed photoelectron, the long range Coulomb potential induces a perturbation on the trajectory dominated by the strong laser field.
Neglecting this Coulomb correction leads to the SFA.

There are a few approaches where the Coulomb correction is taken into account as a first order perturbation along the unperturbed trajectory \cite{Kaushal2013,Landsman2013b}.
At high ellipticity or circular polarization, the Coulomb correction leads to an additional rotation of the final photoelectron momentum in the direction of rotation of the laser field \cite{Landsman2013b}. 

A strong electric field can induce polarization in bound atomic or ionic states and therefore also modify the ion Coulomb potential that the photoelectron feels while propagating in the continuum. But these higher-order terms can often be neglected, provided the $1/r$ Coulomb term is taken into account \cite{Yuan2017}.

This paper is structured as follows. 
Section \ref{sec:Intro} introduced the attoclock method in its originally conceived form, along with all relevant approximations and assumptions. 
Furthermore, alternative experiments were summarized, and more general concepts and approximations of strong field ionization phenomena were presented.

Sections \ref{sec:TDSE} to \ref{sec:StartingTime} build the core of this review. They each discuss recent research and new important developments for the attoclock method.
Section \ref{sec:TDSE} presents an overview of different numerical approaches to the tunnelling time problem.
In section \ref{sec:Dipole} the dipole approximation is investigated.
The single active electron (SAE) approximation, as opposed to taking account of multi-electron effects, is discussed in section \ref{sec:SAEtest}.
In section \ref{sec:2stepModelNA}, non-adiabatic effects and their manifestation in the 2-step model of strong-field ionization are presented. 
Classical trajectory simulations based on the 2-step model are a common tool.
Their details are discussed in section \ref{sec:ClassicalTrajectories}, with special focus on different predictions for the initial conditions probability distribution in phase space at the tunnel exit. 
Finally, section \ref{sec:StartingTime} summarizes work on the starting time of the tunnelling process. 
The paper concludes with section \ref{sec:Summary} summarizing the influences of the different approximations on the interpretation of the attoclock experimental data.

\section{Numerical solutions of the time-dependent Schr\"{o}dinger equation} \label{sec:TDSE}
Since the publication of the first attoclock measurements \cite{Eckle2008,Eckle2008a,Pfeiffer2012}, many groups tried to numerically simulate the experiment by solving the time-dependent Schr\"{o}dinger equation (TDSE)  \cite{Ivanov2014,Torlina2015,Teeny2016,Ni2016,Zimmermann2016,Yuan2017,Sainadh2017,Camus2017,Ni2018a,Bray2018b,Ivanov2018}.
In the case of \cite{Ivanov2014}, the offset angle $\theta$ extracted from the TDSE calculations seem to match with a non-adiabatic field strength calibration of the attoclock experiment data \cite{Boge2013}, see also section \ref{sec:2stepModelNA} and figure \ref{fig:My_Field_Angle_Ivanov2014}.

The authors of \cite{Torlina2015,Sainadh2017,Bray2018b} chose an approach comparing TDSE calculations using a pseudopotential with TDSE results using Yukawa potentials. 
The pseudopotentials are chosen to mimic the screening of $N-1$ bound electrons, such that only a single electron wave function (SAE approximation) is propagated. 
Of course this means that multi-electron effects and polarization of the ion due to the strong field are neglected in these calculations. 
Nevertheless, Yukawa potential calculations where the long-range Coulomb tail is completely suppressed routinely yield negligible streaking offset angles. 
This result is often taken as argument that the observed streaking angle offset $\theta$ of the experiments must be solely due to long-range Coulomb effects \cite{Torlina2015,Sainadh2017,Bray2018b}.
However, one should keep in mind that by replacing the Coulomb potential with a Yukawa potential, either the ionisation potential or the shape of the potential barrier is significantly altered. 

The authors of \cite{Camus2017} commented on this interpretation: ``[...] when the initial nonvanishing momentum of the electron near the tunnel exit is overlooked, the final photoelectron momentum distribution may be explained only with a negative time delay near the tunnel exit.''  Of course, negative tunnelling time would violate causality, illustrating that the choice of initial conditions at the tunnel exit is key to attoclock interpretation.

In \cite{Camus2017}, a quantum mechanical Wigner trajectory \cite{Wigner1955} tracking the most probable photoelectron is calculated, and the results compared to attoclock measurements of argon and krypton. In their analysis, the authors find that a model based on these Wigner trajectories, which includes a finite initial longitudinal momentum at the tunnel exit and finite ionization delay, can reproduce their measurements. 
The issue of the photoelectron momentum at the tunnel exit will be discussed in more details in section \ref{sec:InitLong}.
However, multi-electron effects such as polarization of the ion, or ionization out of a p-orbital rather than an s-orbital, are neglected in this approach, by assuming that these effects are the same for both species, and therefore cancel out when studying the differences between the species \cite{Camus2017}. 

An alternative approach is to monitor the instantaneous ionization rate during the pulse duration \cite{Teeny2016a,Teeny2016,Yuan2017,Yuan2017a} or by applying a tiny signal field \cite{Ivanov2018} and comparing the results to instantaneous tunnel ionization models. 
The probability density current through a virtual detector at the adiabatic tunnel exit point was found to be maximised a finite time $t_i>0$ after the peak of the field \cite{Teeny2016}. 
However, this calculation does not take non-adiabatic effects into account (see section \ref{sec:2stepModelNA} and \cite{Ni2018a}).
In \cite{Yuan2017,Yuan2017a} the authors project their time-dependent wave function onto field free bound states in order to determine the instantaneous ionization rate, finding it lagging behind the peak of the field.
However, this method is not gauge-invariant, contrary to when the projection is executed after the laser pulse has passed \cite{Ivanov2018}.
In the gauge-invariant approach to the instantaneous ionization rate, no asymmetry with respect to the peak of the field strength was found \cite{Ivanov2018}. 
This implies that the tunnelling process is also not asymmetric, meaning that a model assuming starting time $t_0 = 0$ and ionization time $t_i >0$ is not compatible with these results. 
Section \ref{sec:StartingTime} will provide more discussion of the starting time assumption.

Classical backpropagation is yet another TDSE approach \cite{Ni2016,Ni2018,Ni2018a}, exploiting the correspondence of the classical turning point for an electron running up against a potential with the tunnelling exit point. 
In these investigations, the authors defined different exit point criteria for the classical trajectories being propagated backwards in time, after sampling a fully quantum forward calculation. 
They found that if the radial velocity (or the velocity along the instantaneous field direction) should be zero at the exit point, the coordinates are even closer to the ion than in non-adiabatic derivations \cite{Ni2018a,PPT2}. 
The times $t_i$ when these criteria are satisfied are distributed close around the peak of the laser field \cite{Ni2018a}.

The authors of \cite{Zimmermann2016} calculated numerical solutions to the TDSE for strong-field tunnel ionization, and then extracted different tunnelling time predictions defined as derivatives of the complex transmission amplitude \cite{Landsman2015}. Their results show that for this particular approach, the SFA is a good approximation, as long as the field strength does not cross into the over-the-barrier-regime, where the Coulomb potential is suppressed so much that a ground state electron can escape classically.

\section{Dipole Approximation} \label{sec:Dipole}

\begin{figure}[ht]
\centering
\includegraphics[height=0.39\linewidth]{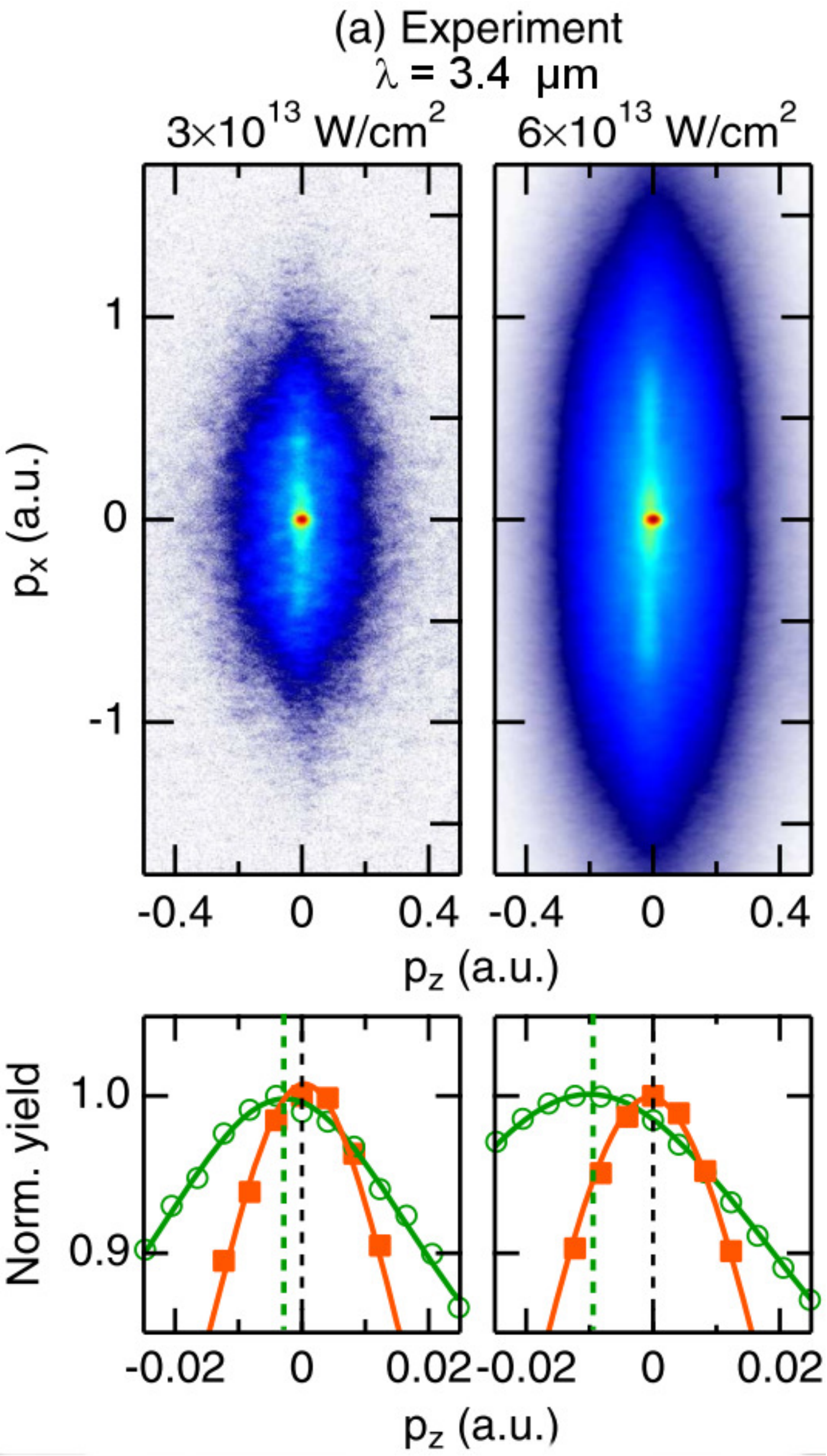}
\includegraphics[height=0.39\linewidth]{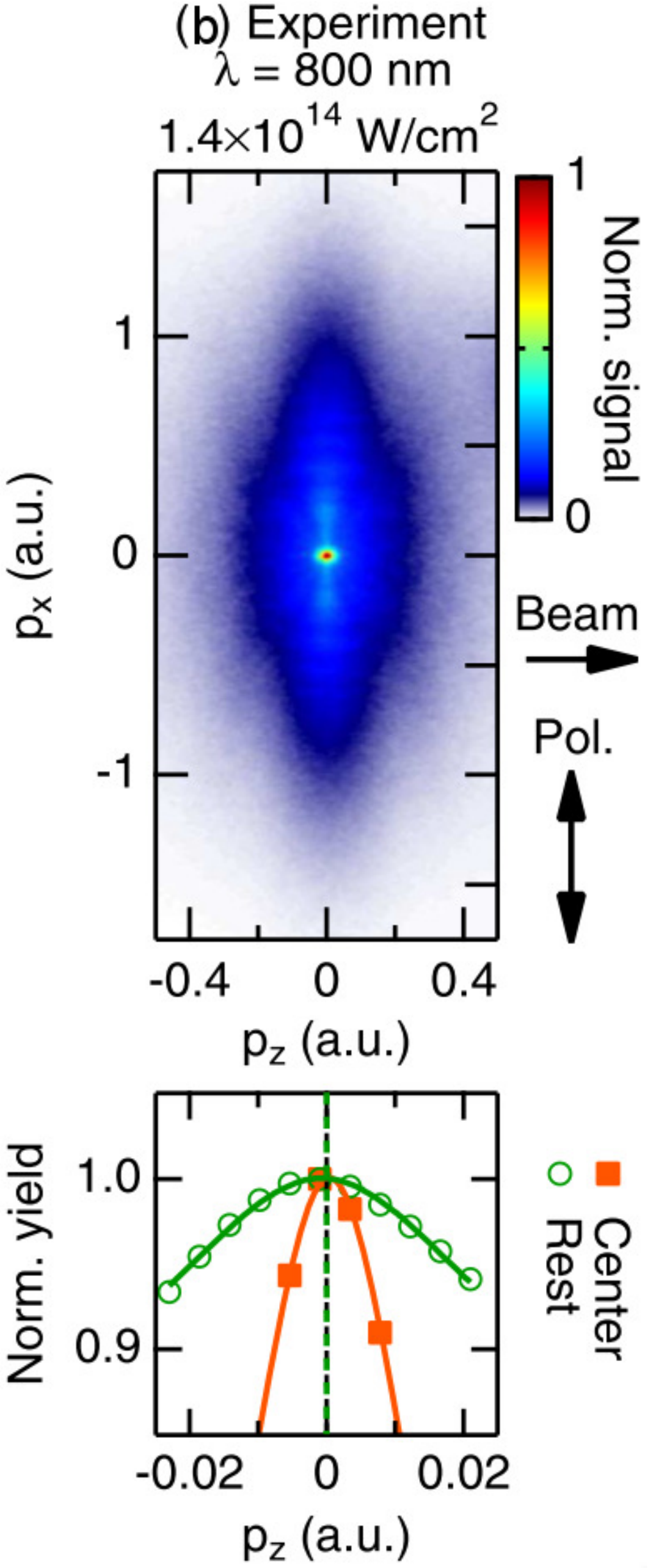} \hfill
\includegraphics[height=0.39\linewidth]{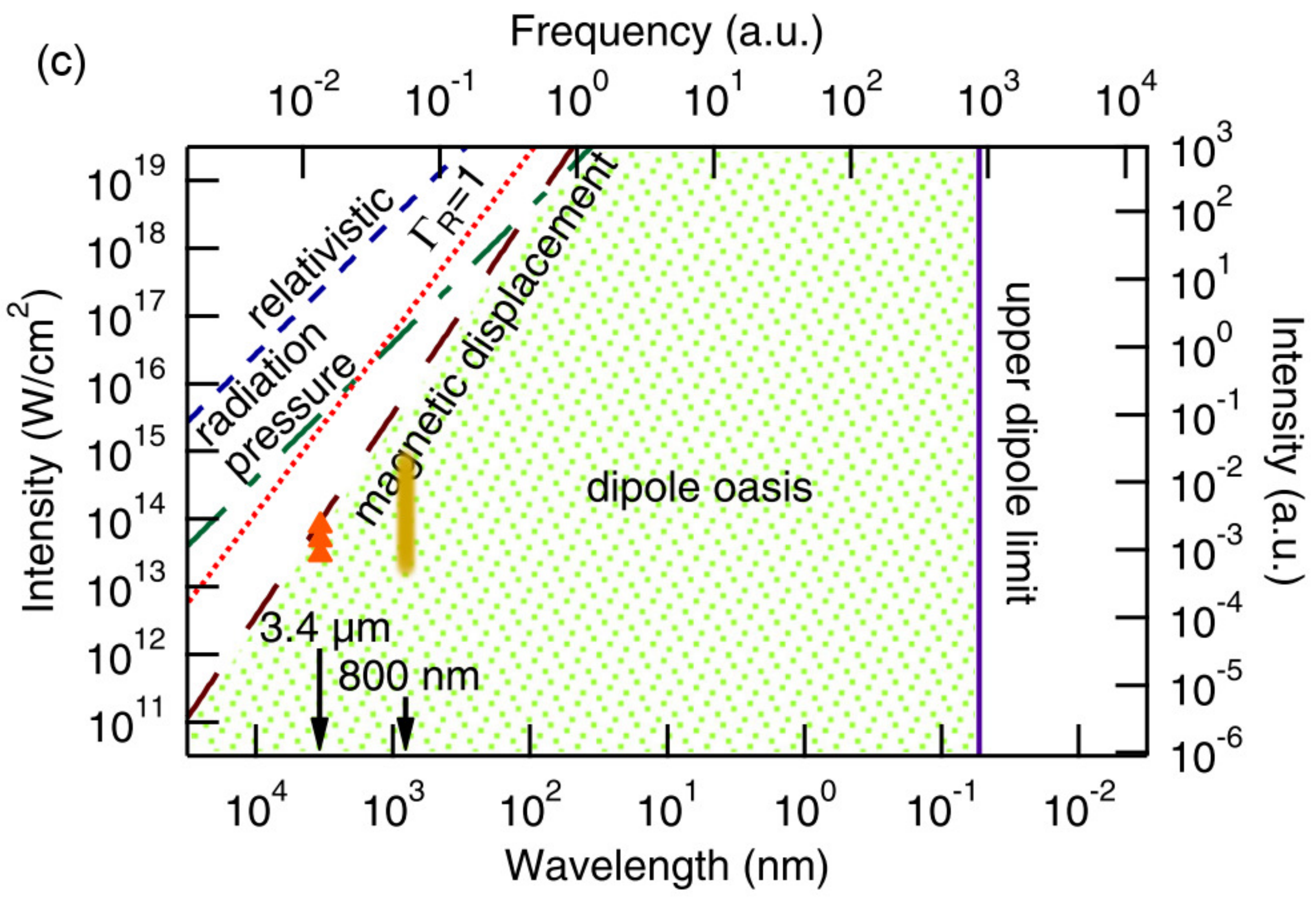}
\caption{The centre dot of the photoelectron momentum distributions (PMD) serves as a reference for absolute zero momentum. The outer PMD ($|p_x|>0.1\,\mathrm{au}$, green circles in histrograms) in panel (a) show a shift in opposite direction to the beam propagation, compared to the centre dot (orange squares) \cite{Ludwig2014,Maurer2018}. The shift can be explained by the onset of magnetic field effects when the laser parameters reach the ``magnetic displacement'' limit of the dipole approximation, see orange triangles in panel (c). Panel (b) shows no such shift for laser parameters as they were used in the attoclock experiment, see yellow area in panel (c). Figures adapted from \cite{Ludwig2014}. }
\label{fig:Ludwig_fig}
\end{figure}

The dipole approximation is easily satisfied in the experimental cases studied in the attocklock experiment, and related calculations. The near-IR field of 735 nm at intensities of 0.3 up to $8\cdot 10^{14}\,\mathrm{W/cm^2}$ is a regime well within both limits, see figure \ref{fig:Ludwig_fig}(c). 
The wavelength is long enough that the photoelectrons do not feel the spatial dependence, and the influence of the magnetic field is negligibly small. 
To illustrate this, the authors of \cite{Ludwig2014,Danek2018} performed photoelectron momentum measurements in linear polarization for $\lambda = 3.4\,\mathrm{\mu m}$ as well as $\lambda = 800 \, \mathrm{nm}$. As can be seen from figure \ref{fig:Ludwig_fig}(a), the effect of the magnetic field causing a shift of the photoelectron momenta opposite the beam propagation direction is only visible when the experimental parameters reach beyond the ``dipole oasis''.
The same effect is absent for experiments within both the upper and lower wavelength limit, which is the case for the attoclock measurements, compare figure \ref{fig:Ludwig_fig}(b).

\section{Single Active Electron vs. Multi-electron effects} \label{sec:SAEtest}

In semiclassical and quantum mechanical treatment of strong-field ionization, it is common to use the SAE approximation, assuming that only one valence electron will tunnel ionize, while the rest of the bound electrons end up in ionic ground state. This of course invites questions on the validity of any model based on the SAE when interpreting experimental results for multi-electron atoms or molecules. 

\begin{figure}[ht]
\centering
\includegraphics[width=0.7\linewidth]{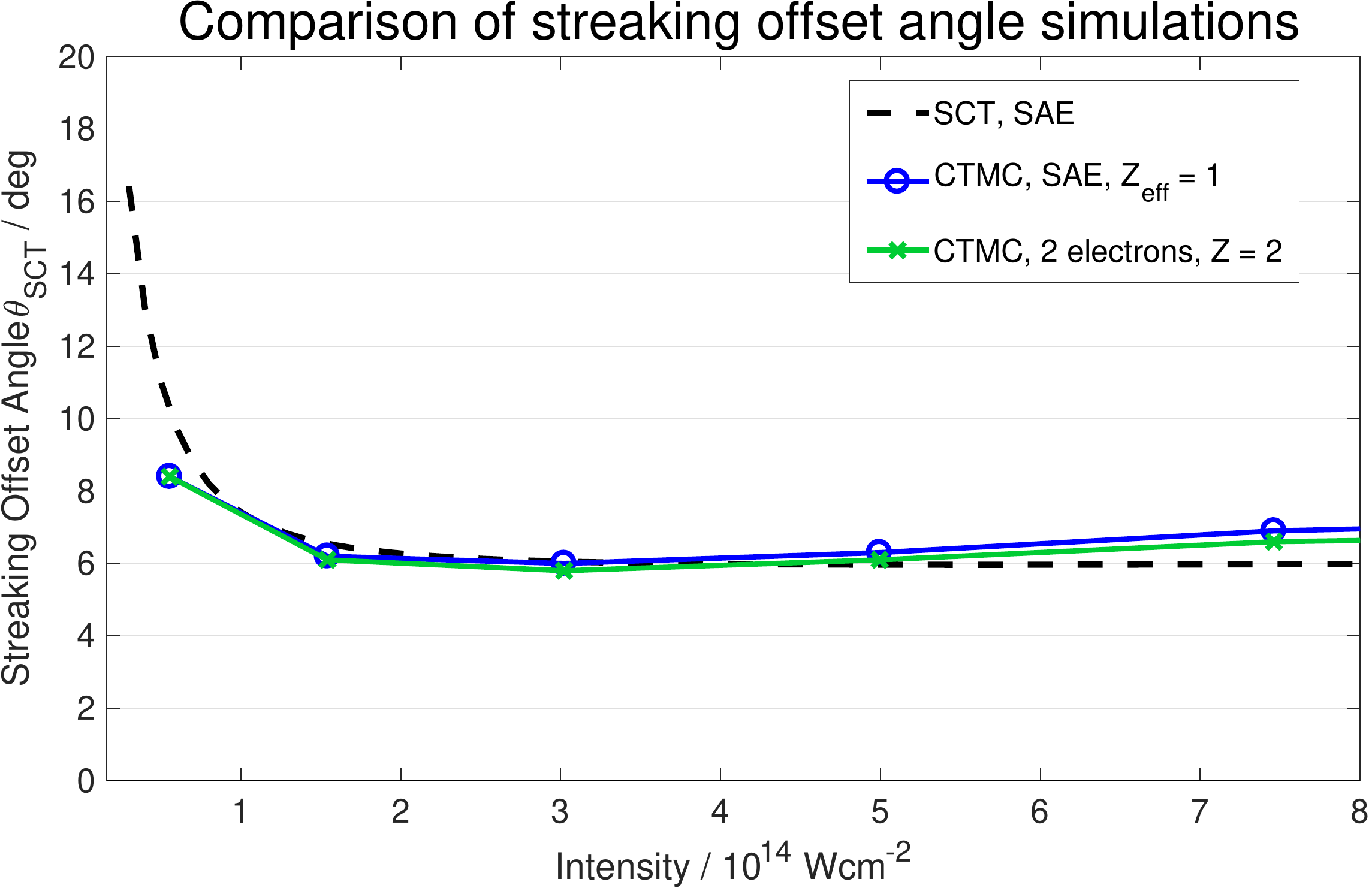} 
\caption{\textbf{Streaking offset angles $\theta_{\mathrm{SCT}}$ comparison between three numerical models}: The single active electron (SAE) single classical trajectory (SCT) calculation shown in black dashed line was used in the analysis of the attoclock measurement \cite{Landsman2014b,Landsman2014}. 
Classical trajectory Monte Carlo (CTMC) calculations were computed using an independent code \cite{Emmanouilidou2015}, once with the SAE approximation (blue solid line with $\ocircle$), and once as a two-electron (three-body) calculation (green solid line with $\times$). 
Calculations based on the SAE approximation agree with the calculation including both electron-electron interaction as well as electron-nuclear force. 
All calculations shown in this figure assume an adiabatic framework. 
Figure adapted from \cite{Emmanouilidou2015}.}
\label{fig:Experiment}
\end{figure}

The exchange and interaction between an ionized and a second bound electron in helium was studied with CTMC methods, focusing on the post-ionization dynamics \cite{Emmanouilidou2015}. 
It was found that an effective Coulomb potential with $Z = 1$, corresponding to perfect screening by the remaining bound electron(s), reproduced the final photoelectron momentum distribution (PMD) of two-electron calculations, see figure \ref{fig:Experiment}. 
Therefore, it is safe to neglect multi-electron effects during the continuum propagation of the ionized electron. 

Even if multi-electron effects are negligible once the ionized electron is already far away from the parent ion, there still might be significant electron-electron interaction during the actual tunnel ionization step, while the tunnelling electron is still at a comparable distance to the nucleus relative to the other bound electron. 
A similar analysis for the tunnel ionization step, however, is challenging to perform since it requires a fully quantum mechanical treatment. Near-circular, but not perfectly circular, polarization prohibits coordinate reduction based on symmetry arguments, making the numerical solution of the TDSE computationally very expensive. 
Recently, Majety and Scrinzi \cite{Majety2017} published an approach for reducing the necessary basis functions with higher orbital angular momentum. 
The results of \cite{Majety2017} show that, similar to the propagation in the continuum, the tunnelling step can also be approximated with a single active electron for the case of helium. They could not find any observable differences in the final angular momentum spectrum between a SAE calculation and multi-channel calculations \cite{Majety2017}. 

Both these studies leave us with the conclusion that the single active electron approximation is valid, at the very least for helium atoms as the target and the laser parameter range studied in the attoclock experiment.
For larger atoms, there is less prior work focusing on this aspect. 
Though there is evidence that some multi-electron effects, specifically the polarization of the remaining parent ion in the strong laser field, can significantly influence the trajectory of the ionized photoelectron for the case of argon \cite{Shvetsov-Shilovski2012,Pfeiffer2012}.
Even more so, for tunnel ionization of molecules the polarizability and electron-electron interactions are important to take note of \cite{Lezius2001}.

\section{2-Step model with non-adiabatic effects} \label{sec:2stepModelNA}

The original attoclock experiment was evaluated in the adiabatic approximation \cite{Eckle2008a,Landsman2014b} characterized by a Keldysh parameter \cite{Keldysh1965} of 
\begin{equation}
\gamma := \frac{\omega\sqrt{2 I_p}}{F} \ll 1,
\end{equation}
as is typical for strong-field experiments in a similar intensity range \cite{Shafir2012,Hickstein2012,Odenweller2014,Wolter2015,Landsman2013,Landsman2013b}. 
However, non-adiabatic effects influence especially the field strength calibration of strong-field ionization data already significantly for $\gamma \approx 1$
\cite{Boge2013,Hofmann2014,Hofmann2016,Cai2017}. 
This calls for a thorough reevaluation of the original attoclock data interpretation.

Taking account of the dynamics of the strong electric field leads to several effects which are neglected in the adiabatic approximation.
During the tunnelling process, the electron wave packet can gain energy from the oscillating field. This results in a shorter tunnel exit radius of the photoelectron compared to the quasistatic estimate \cite{Mur2001,Ni2016}, compare also figure 3 of \cite{Landsman2015}.
Also, the ionization probability falls off slower with reducing field strength compared to the adiabatic prediction (see figure 2 in \cite{Yudin2001}), and the PMDs are predicted to be wider in the non-adiabatic case than in the adiabatic approximation.
Furthermore, for the case of elliptical or circular polarization, the rotation of the field is imprinted onto the photoelectron, which exhibits an initial transverse momentum tangential to the rotation of the electric field at the tunnel exit \cite{PPT2}. 
This initial transverse momentum in turn yields a larger final absolute momentum for the same field strength compared to the adiabatic formalism, which strongly influences the field strength calibration of experimental data at lower intensities \cite{Hofmann2014,Hofmann2016,Cai2017}.
For experimental data, the field strength which the photoelectrons experienced must be calibrated a posteriori from the measured PMD, by comparing a measured observable to predictions from a model \cite{Hofmann2016}. 
This leads to a shift of the (same) experimental data to lower field strengths if treated in the non-adiabatic framework.

The same experimental data of \cite{Landsman2014b} has already been studied in another publication in order to assess non-adiabatic effects \cite{Boge2013}. 
The authors of \cite{Boge2013} focused on the influence of the initial transverse momentum on the angle of the most probable final momentum.
On the other hand, for the calculation of SCT and CTMC simulations, the shorter exit radius in the non-adiabatic framework was neglected in this particular work. 
The choice of such a mixed adiabatic/non-adiabatic model lead the authors to conclude that the attoclock data does not exhibit non-adiabatic effects.
This conclusion was questioned by later work, where fully non-adiabatic models were considered \cite{Ivanov2014,Hofmann2014,Hofmann2016,Cai2017}.

Two other works \cite{Ivanov2014,Klaiber2015} looked at the original attoclock data in connection with non-adiabatic effects. 
The first calculated the numerical solution to the TDSE for a small range of intensities covered in the experiment \cite{Ivanov2014}.
The second used an analytical model based on the standard SFA methodology \cite{Anatomy}, but extending it to explicitly include non-adiabatic dynamics as well as influence of the Coulomb potential during the tunnelling process and the propagation in the continuum \cite{Klaiber2015}.

Furthermore, the authors of \cite{Cai2017} combined ideas of \cite{Boge2013} and \cite{Hofmann2016} to check for non-adiabatic effects with TDSE calculation as well as directly in the attoclock offset angle measurements. They also concluded that non-adiabatic effects must be taken into account, and that the sub-barrier quantum motion is important and should not be neglected in strong-field ionization models \cite{Cai2017}.  

\begin{figure}[ht]
\centering
\includegraphics[width=0.8\linewidth]{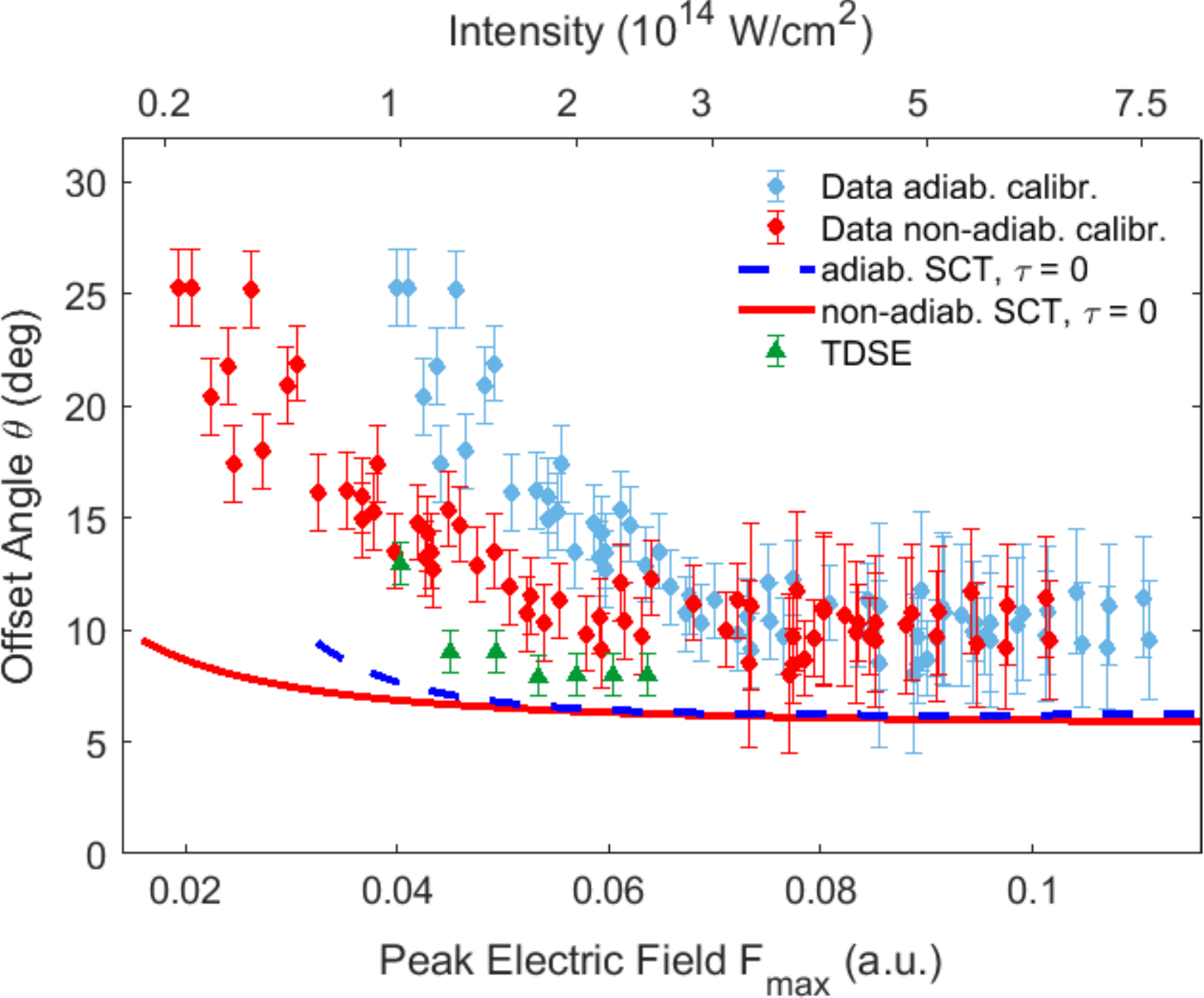}
\caption{\textbf{Effect of field strength calibration:} Comparison of measured streaking offset angles with single classical trajectory (SCT) reference calculations assuming instantaneous tunnelling ($\tau = 0$). 
The red solid line shows the prediction by non-adiabatic SCT simulation, while the blue dashed line represents the adiabatic prediction.
For the case of helium, the adiabatic and non-adiabatic SCT yield the same angle prediction for a large range of field strengths. 
Also shown are the values extracted form TDSE by \cite{Ivanov2014} as green triangles. }
\label{fig:My_Field_Angle_Ivanov2014}
\end{figure}
Figure \ref{fig:My_Field_Angle_Ivanov2014} shows the attoclock data \cite{Boge2013,Landsman2014b} in adiabatic and non-adiabatic calibration (blue and red dots), compared to the TDSE calculation for the final streaking angle by \cite{Ivanov2014} (green triangles).
Evidently, the calculation agrees with the non-adiabatic calibration of the measurement data.
Additionally, SCT calculations of the expected streaking offset angle $\theta_{\mathrm{SCT}}$ for instantaneous tunnelling are shown as blue dashed line for the adiabatic approximation, and red solid line including non-adiabatic effects. 

\section{Classical trajectories} \label{sec:ClassicalTrajectories}

In attoclock experiments, the experimental observable (offset angle $\theta$) is compared to a zero-time-calibration calculated within a model assuming instantaneous tunnelling, where the angle is typically computed using Classical Trajectory Monte Carlo (CTMC) simulations. 
The computational costs for CTMC simulations is very low compared to quantum simulations. 
Therefore, CTMC simulations can achieve highly precise converged results.
This gives them a distinct advantage over analytic approaches, which may only be applicable over a narrow range of conditions, such as in \cite{Bray2018b} (see next paragraph for detail).
The driving laser field, the Coulomb potential of the residual ion, dipole effects in the ion due to induced polarization by the laser field, and even electron-electron correlation \cite{Emmanouilidou2015} can all be fully and explicitly taken into account. 
Of course, the accuracy of the resulting calibration hinges on the distribution function and the sampling of the initial conditions. 
The analytical probability distribution functions in phase space must accurately describe the ionized part of the wave function after a quantum tunnel ionization process.

Recently, Bray, Eckart and Kheifets suggested an analytic approach that neglects the laser field during propagation, estimating the Coulomb correction using  Rutherford scattering angle in an attractive potential \cite{Bray2018b}.
This so-called Keldysh-Rutherford (KR) model uses the adiabatic approximation which neglects the energy gain during the tunnelling process and the initial transverse momentum of the photoelectron at the tunnel exit,
although these non-adiabatic effects are increasingly prominent for low intensities. 
The scattering parameter $\rho$ is assumed to be the same as the exit radius, although formally $r_{\mathrm{exit}} < \rho$, unless the energy of the scattering particle is infinite. 


Since the Rutherford formula gives the scattering angle in the absence of any time-dependent fields, the KR formula becomes increasingly accurate when the laser field has less of an impact, meaning for weaker intensities and shorter pulses.  Hence, it may not be applicable to any existing attoclock experimental data, which requires sufficiently strong laser fields to achieve tunnel ionization.  

It may nevertheless be instructive to apply the KR formula to the recent experimental data on hydrogen, as suggested by the authors in \cite{Bray2018b}:   ``Because of its
simplicity, the Keldysh - Rutherford formula can be easily applied to
attoclock experiments with arbitrary polarization though modification
of the above formalism to account for nonunitary ellipticity. One such
case being the recent attoclock measurements on atomic hydrogen \cite{Sainadh2017},
where the signature field intensity scaling of the KR model $I^{0.5}$ was
indeed observed."

\begin{figure}[h]
\centering
\includegraphics[width=0.7\linewidth]{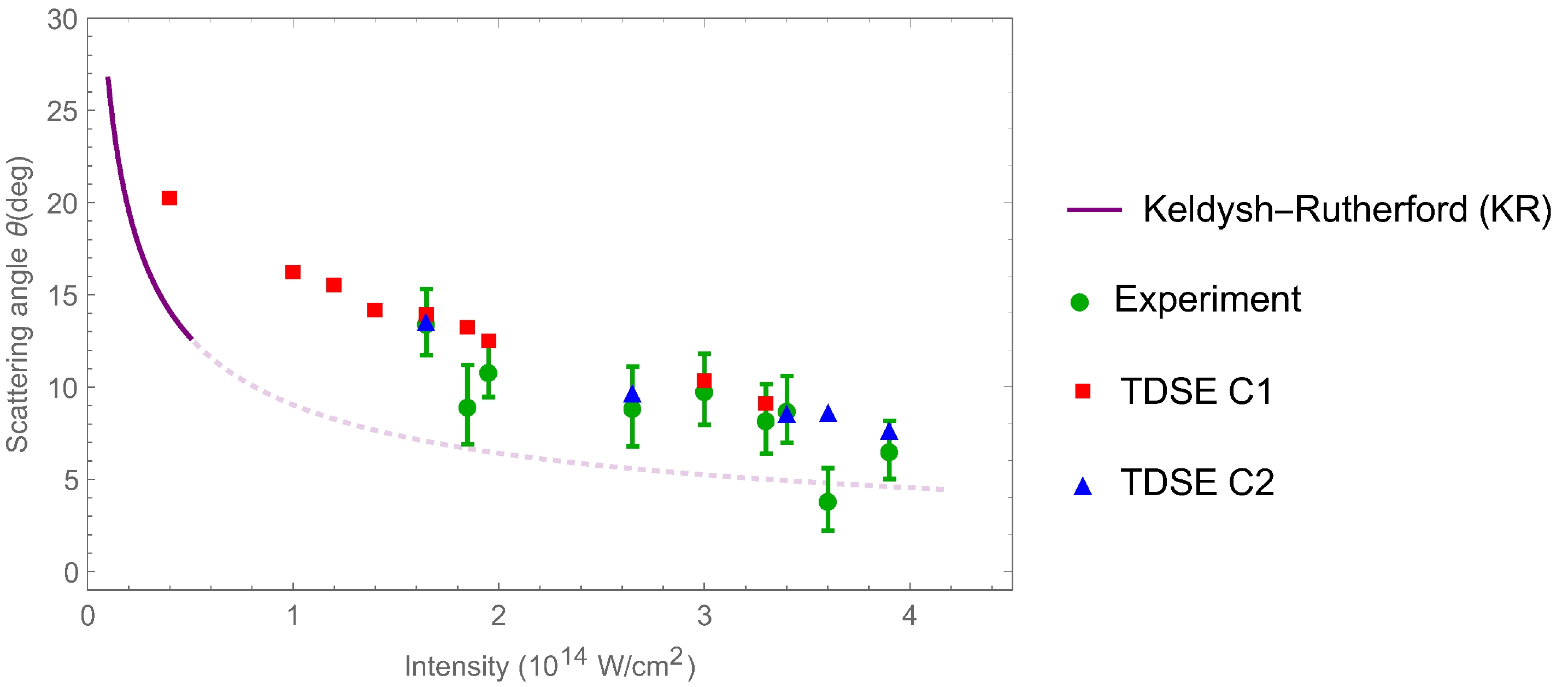} 
\caption{Applying the Keldysh-Rutherford model (KR) \cite{Bray2018b} to the attoclock experiment on atomic hydrogen by \cite{Sainadh2017}. Red squares and blue triangles show the offset angle $\theta$ extracted from two different TDSE calculations \cite{Sainadh2017}, while the green dots with error bars are the experimental values \cite{Sainadh2017}. The KR model predicts that the offset angle due to Coulomb scattering is smaller than the measured or calculated total offset angle, suggesting significant tunnelling time. Note however that the KR model may be inapplicable to existing strong field ionization experiments (see text for detail). 
	}
\label{fig:RutherfordVSainadh}
\end{figure}

Following the above quote, we plotted the KR formula alongside attoclock measurements on atomic hydrogen \cite{Sainadh2017}.  The results are shown in figure \ref{fig:RutherfordVSainadh}, alongside with TDSE simulations also presented in \cite{Sainadh2017}.  As figure \ref{fig:RutherfordVSainadh} illustrates, there remains an angle difference between the KR estimate, the TDSE calculations and the experimental data, suggesting non-negligible tunnelling time.  

\subsection{General CTMC and SCT} \label{sec:SCTvsCTMC}
The classical trajectories for the attoclock configuration start at an exit radius of approximately 8 au or larger from the ion core \cite{Landsman2015}. 
Due to the elliptical polarization of the field, which creates a transverse drift in electron momentum, these trajectories typically never return to the vicinity of the parent ion. 
Due to the weak influence of the Coulomb potential after ionization (particularly in the case of elliptically polarized light), and the absence of resonances or other strong phase shifts (compare section \ref{sec:SPI}), the quantum-classical correspondence is valid \cite{Ehrenfest1927}. 

\begin{figure}
\centering
\includegraphics[width=\linewidth]{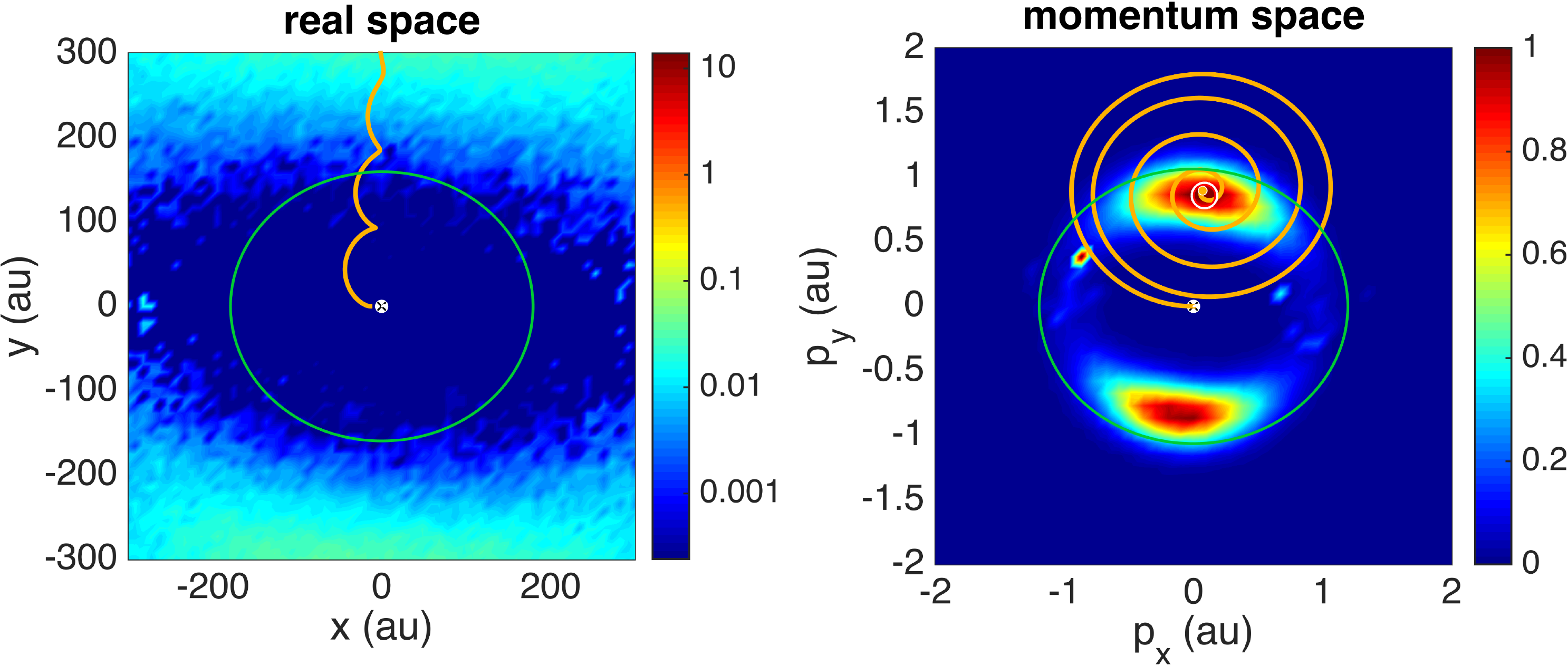} 
\caption{\textbf{Most probable trajectory} The colour scale shows the classical trajectory Monte Carlo (CTMC) simulation real space (left) or momentum space (right) probability density after the laser pulse has passed. The orange line traces a single classical trajectory (SCT). The target was helium, irradiated by a laser field with the following parameters: $\epsilon=0.89$ (indicated as the green solid polarization ellipse), $\lambda = 735\,\mathrm{nm}$, pulse duration FWHM $9\,\mathrm{fs}, I = 2.5\cdot 10^{14}\,\mathrm{W/cm^2}$. The influence of the ion Coulomb force on the electron during the propagation is included.
A SCT initiated with the most probable initial conditions traces the highest probability dencity of the wave packet. See suplemental material for a movie version \cite{Supp}.}
\label{fig:CTMCdensityframe0300}
\end{figure}

Additionally, a single classical trajectory (SCT) launched with the most probable initial conditions follows the propagation of the highest probability density in the full CTMC simulation, see figure \ref{fig:CTMCdensityframe0300}.
More details on our implementation of CTMC simulations based on adiabatic Ammosov, Delone \& Krainov (ADK) models \cite{ADK1986,Delone1991} can be found in \cite{Hofmann2013,Hofmann2016thesis}.

\subsection{Implementation of non-adiabatic effects} \label{sec:NA_CTMC}

The most popular non-adiabatic strong-field ionization theory was developed by Perelomov, Popov and Terent'ev (PPT) \cite{PPT1,PPT2}, and rewritten as \cite{Mur2001}. 
This analytical approach describes the final photoelectron momentum probability distribution averaged over one laser cycle, for arbitrary ellipticity of the ionizing field. 
Non-adiabatic models deriving an instantaneous ionization rate $\Gamma(t)$ are typically only valid for linear (and sometimes circular) polarization \cite{Yudin2001,Bondar2008,Li2016}. 
In order to describe classical trajectories starting at different times during the laser pulse, we introduced the time dependence by letting the Keldysh parameter $\gamma$ depend on the instantaneous field strength $|F(t)|$. 
The energy gain of the photoelectron during the tunnelling process results in a shorter exit radius compared to the adiabatic version, and the initial transverse momenta follow a Gaussian distribution centred about the most probable initial transverse momentum.
For more details on the non-adiabatic CTMC implementation, please refer to \cite{Hofmann2014,Hofmann2016thesis}.
Table \ref{tab:CTMCoverview} compares the main characteristics of the CTMC simulations concerning both the sampling of initial conditions and the classical propagation.
\begin{table}[h]
	\caption{Overview of different characteristics of the different classical trajectory simulations, based on either adiabatic Ammosov, Delone \& Krainov (ADK) \cite{ADK1986,Delone1991} theory, or non-adiabatic Perelomov, Popov \& Terent'ev (PPT) \cite{PPT1,PPT2,Mur2001} theory.} 
	\raisebox{-0.5\height}{\includegraphics[width=0.45\linewidth]{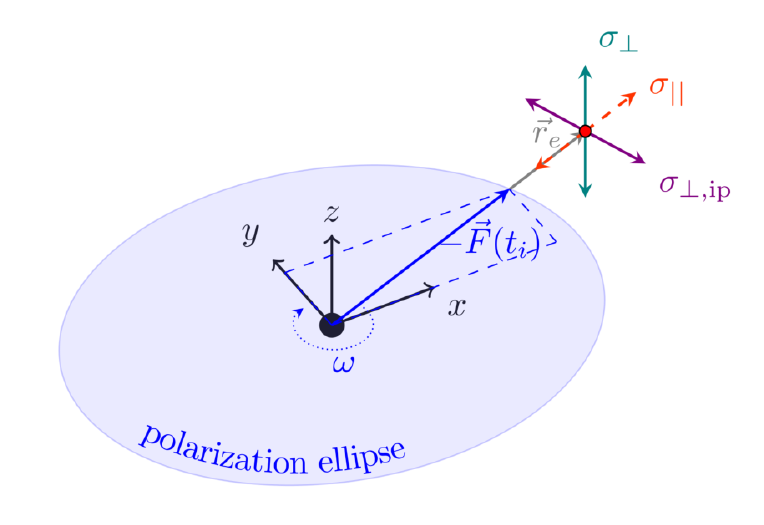}}
	\begin{tabularx}{0.48\textwidth}{X}
		{\footnotesize The figure illustrates the definition of longitundial ($||$) and orthogonal ($\perp$) momentum components, relative to the instantaneous field direction at the starting time $t_i$ for a classical trajectory.}
	\end{tabularx}
	\label{tab:CTMCoverview}
\begin{tabular}{lll}
	\toprule
	characteristic	& adiabatic CTMC \cite{Hofmann2013} & non-adiabatic CTMC \cite{Hofmann2014} \\
	\midrule
	\multicolumn{3}{l}{\emph{starting conditions:}} \\
	$\Gamma(t)$	&	exponential ADK	&	PPT, with modified $\gamma(t)$ \\
	$\mathbf{p}_{\perp}^i$	and $\sigma_{\perp}$, $\sigma_{\perp,\mathrm{ip}}$&	ADK	& PPT \\
	$p_{||}^i$ 	and $\sigma_{||}$& 0	&	0 \\
	$r_{e}$	&	parabolic coordinates \cite{Landau1965,Fu2001,Shvetsov-Shilovski2012}	&	PPT \\
	$\IP$	& Stark shift included \cite{Shvetsov-Shilovski2012}	& Stark shift included \\
	\midrule
	\multicolumn{3}{l}{\emph{propagation for both adiabatic and non-adiabatic case:}} \\
	ion Coulomb:	& \multicolumn{2}{l}{soft-core potential: $V(r) = \frac{-1}{\sqrt{r^2+a}}$,  with $a = 0.1 \,\mathrm{au}^2$}  \\
	induced dipole:	& \multicolumn{2}{l}{same soft-core constant $a$} \\
	electric field:	& \multicolumn{2}{l}{always included using dipole approximation} \\
	bound electrons:	& \multicolumn{2}{l}{single active electron approximation, $Z_{\mathrm{eff}}=1$} \\
	\bottomrule
\end{tabular}
\end{table}

Figure \ref{fig:My_Field_Angle_Ivanov2014} demonstrates the difference in the field strength calibration of the measured data, where the blue dots are the values from \cite{Landsman2014b}, and the red dots are recalibrated based on the PPT theory \cite{PPT2}.
The red solid line in figure \ref{fig:My_Field_Angle_Ivanov2014} shows the non-adiabatic PPT\cite{PPT2} and the blue dashed line the adiabatic 
\cite{Pfeiffer2012,Shvetsov-Shilovski2012} prediction of the streaking offset angle.
These SCT simulations yield the Coulomb correction on the field induced streaking angle assuming instantaneous tunnelling.
For the case of helium, the two non-adiabatic effects of initial longitudinal momentum and shorter exit radius seem to cancel each other out over a large range of field strength, essentially predicting the same final streaking angle offset as the adiabatic approximation. 
Within the attoclock framework, the angle difference between the measurements and the zero-time reference calculation SCT is then interpreted as being due to a delayed release of the electron into the continuum.
Evidently, taking non-adiabatic effects into account still results in a significant streaking angle difference between what is measured and what is expected under the assumption of instantaneous tunnelling.

\subsection{Influence of initial longitudinal momentum distribution} \label{sec:InitLong}

A core approximation in many analytical descriptions of strong-field ionization is zero momentum of the photoelectron at the tunnel exit parallel to the direction of the electric field (longitudinal), $p_{||}^i=0$ \cite{Anatomy,PPT2,Ni2016}.
However, there are several independent works suggesting that the initial longitudinal momentum should be a spread \cite{Wavepacket,Hofmann2013,Hofmann2014,Bondar2008}, and possibly even with a non-zero most probable value \cite{Teeny2016,Camus2017}. 
Ni \etal found complementary results with their classical backpropagation method. 
The classical turning point, when the photoelectron has zero momentum parallel to the electric field, was located at a position even closer to the ion than what PPT predicts \cite{Ni2016,Ni2018a}. 
This could intuitively be understood as the photoelectron having gained some outwards momentum already by the time it passes by the exit radius predicted by PPT.

\begin{figure}[ht]
	\centering
	\includegraphics[trim=0cm 0cm 3cm 0cm,clip=true,width=0.42\linewidth]{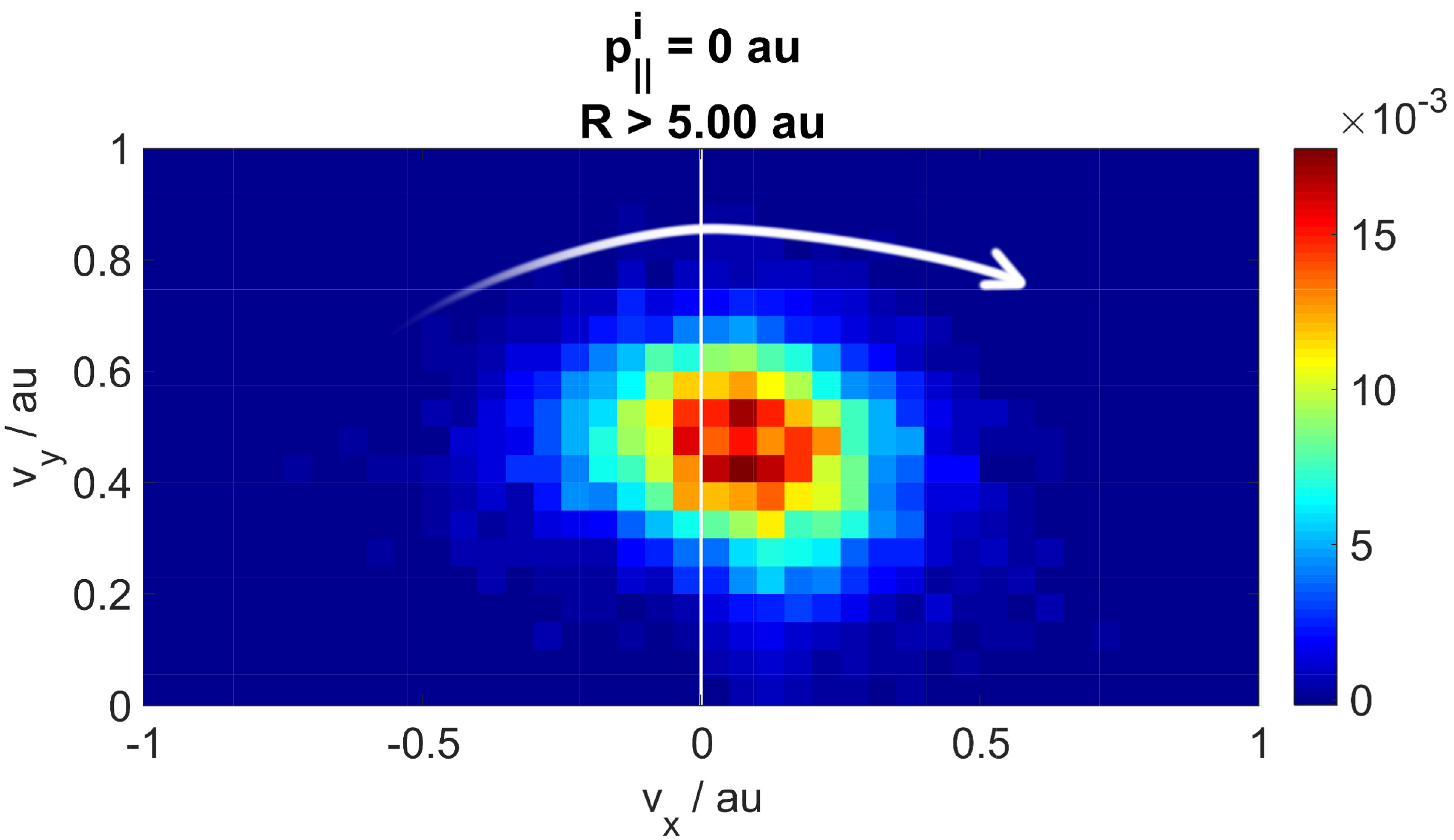} 
	\includegraphics[trim=3cm 0cm 0cm 0cm,clip=true,width=0.42\linewidth]{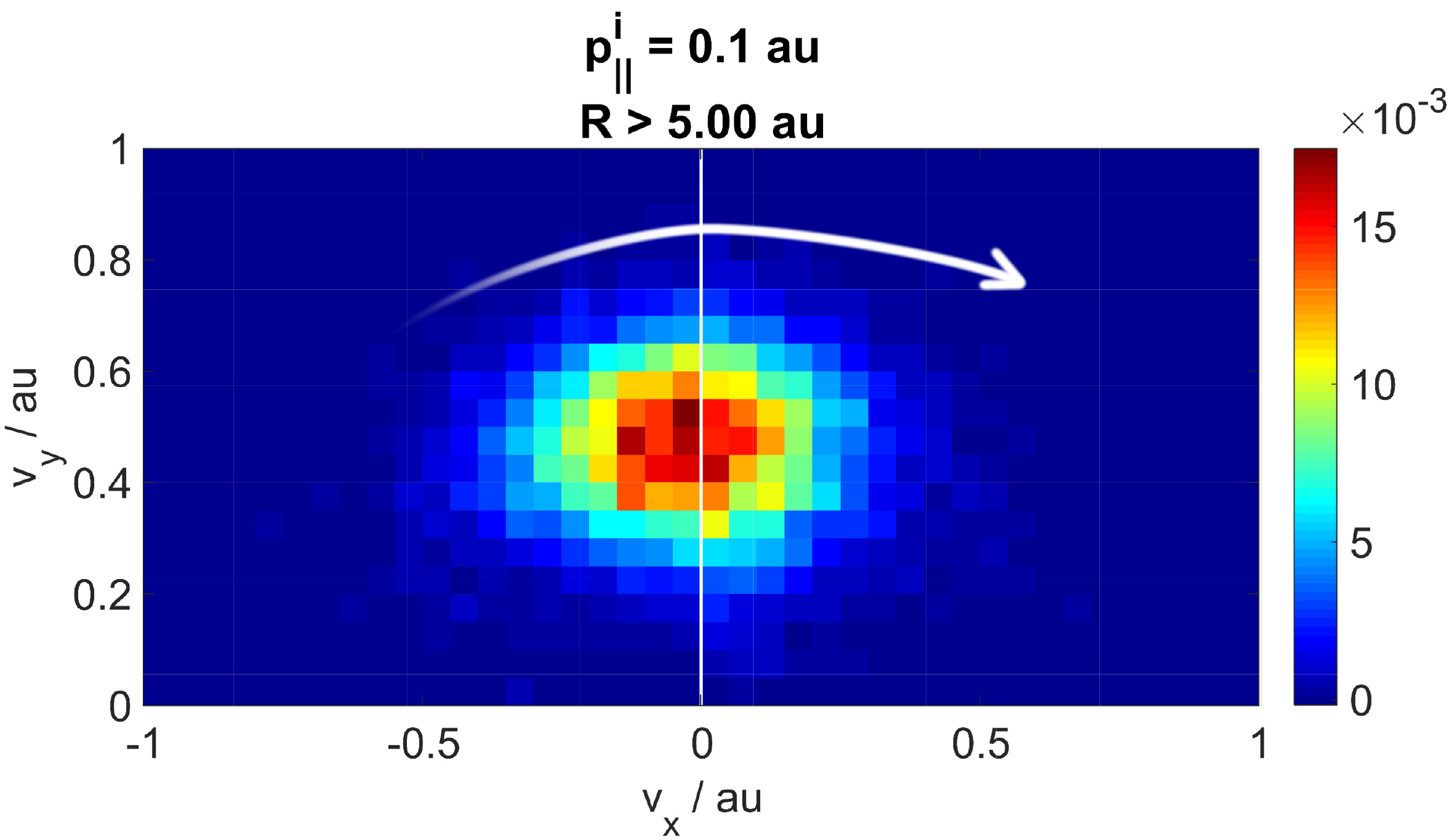} 
	\caption{Final photoelectron momentum distribution (PMD) calculated by adiabatic classical trajectory Monte Carlo (CTMC) simulations in the $v_y>0$ half-plane. The two panels compare a CTMC with zero (left) or finite (right) most probable initial longitudinal momentum $\mathbf{p}_{||}^i$. The laser field is rotating clockwise, as indicated by the white arrows. The figures illustrate that the offset angle $\theta_{\mathrm{SCT}}$ becomes smaller when $\mathbf{p}_{||}^i>0$ is assumed. This would lead to a larger angle difference $\theta - \theta_{\mathrm{SCT}}$ between the measured offset angle $\theta$ and the zero-time calibration $\theta_{\mathrm{SCT}}$ (compare also figure \ref{fig:VMIdataSCT}). }
	\label{fig:e87_s0_EndVelocitiesDRaC}
\end{figure}

Taking account of a positive most probable initial longitudinal momentum leads to a \emph{reduction} of the SCT prediction for the final streaking angle $\theta_{\mathrm{SCT}}$. 
Based on the conservation of canonical momentum, the final momentum is shifted by $\mathbf{p}_{||}^i$ compared to the simulations assuming zero initial longitudinal momentum.
This effect is visible in figure \ref{fig:e87_s0_EndVelocitiesDRaC}, where the panel on the right includes a non-zero initial longitudinal momentum, and the white arrow denotes the rotation sense of the driving field. 

\begin{figure}
\centering
\raisebox{-0.5\height}{\includegraphics[width=0.45\linewidth]{figure10a-eps-converted-to}} \hfill 
\raisebox{-0.5\height}{\includegraphics[width=0.54\linewidth]{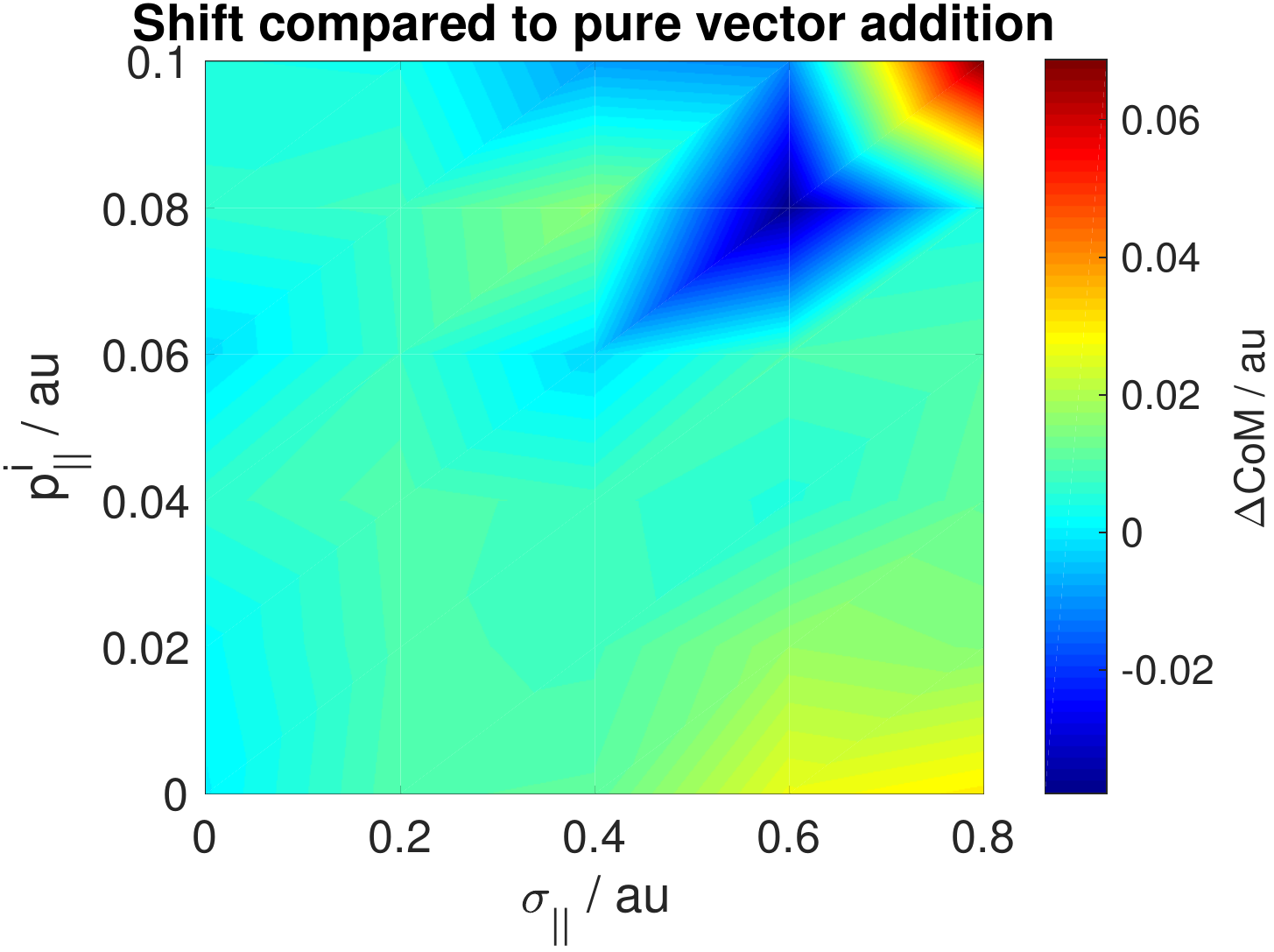}} 
\caption{\textbf{Coulomb deformation due to an initial longitudinal momentum:} The colour scale represents the deviation $\Delta \mathrm{CoM}$ of the centre of mass of the final photoelectron momentum distribution (PMD) compared to an expected shift of $\mathbf{p}_{||}^i$ based on the vector addition of the conservation of momentum. The values were extracted from adiabatic classical trajectory Monte Carlo simulations. The majority of the tested range of initial most probable momentum $\mathbf{p}_{||}^i$ and initial longitudinal momentum spread $\sigma_{||}$ only exhibits very small deviations for the CoM away from the pure vector addition. The sketch on the left again shows the coordinate definitions (adapted from \cite{Hofmann2016}).}
\label{fig:DeltaCoM}
\end{figure}

This observation leads to a related question. Does the influence of the Coulomb force result in an asymmetric deformation of the photoelectron wave packet?
For figure \ref{fig:DeltaCoM}, the centres of mass (CoM) of CTMC calculations with varying spread $\sigma_{||}$ and most probable value $\mathbf{p}_{||}^i$ for the initial longitudinal momentum distribution at the tunnel exit are extracted. Their values are then compared to the naive expectation of 
\begin{equation}
\mathrm{CoM}^{\mathrm{expected}} = \mathrm{CoM}(\sigma_{||}=0, \mathbf{p}_{||}^i=0) + \mathbf{p}_{||}^i,
\end{equation}
which is based on simple vector addition within the conservation of canonical momentum. 
The difference between the actually extracted CoM and this expected value is plotted in figure \ref{fig:DeltaCoM} (colourmap). All determined shift-differences are smaller than one bin size of the PMD in figure \ref{fig:e87_s0_EndVelocitiesDRaC}, and thus negligible, with the sole exception of the extreme case of $\sigma_{||}=0.8\,\mathrm{au}, \mathbf{p}_{||}^i=0.1\,\mathrm{au}$.
Therefore, we can conclude that the asymmetric influence of the Coulomb force is negligibly small, and SCT are still a valid and easy approach to determine the classical trajectory prediction for the most probable final momentum.

\section{Starting time assumption} \label{sec:StartingTime}

For determining the duration of the tunnel ionization process, knowing the moment of when an electron exits from the potential barrier is of course not sufficient. 
The starting point $t_0$, when an electron enters the potential barrier (in a pseudo-classical picture) must be defined, too. 
In strong-field ionization models such as PPT \cite{PPT1,PPT2}, ADK \cite{ADK1986,Delone1991,Anatomy}, or many others \cite{Yudin2001,Bondar2008,Klaiber2015,Li2016,Li2017} this intuitive definition is typically assigned to the complex transition point $t_s$, which is a time calculated by the saddle point approximation \cite{Anatomy}. 
However, most of these models, with the notable exception of \cite{Klaiber2015}, then either define the ionization time $t_i$ to be the real part of $t_s$, or the calculation automatically yields that relation due to a short-range potential approximation, and neglecting non-adiabatic effects during the tunnelling process \cite{Anatomy,Klaiber2015}. 
This then leads to the interpretation that there is no (real) time passing while the electron tunnels through the potential barrier, since 
\begin{equation}
\tau = t_i - t_0 = \Re{t_s} - \Re{t_s} = 0.
\end{equation}

There are several publications suggesting that the starting time should be before the ionizing field reaches its maximum. 
In \cite{Teeny2016,Teeny2016a}, the authors monitor the probability current density in a one-dimensional TDSE calculation of strong field ionization.
At the classical tunnel entry point $x_{\mathrm{in}}$, they find that the outflowing current maximizes clearly before the electric field reaches its maximum value. 
Furthermore, for a large range of intensities tested in \cite{Ni2016}, the classical backpropagation (after two-dimensional TDSE forward calculation) reaches classical turning points at times $t_i \lesssim 0$.
By causality therefore, the starting times also must be $t_0 \lesssim 0$. 
In the Coulomb-corrected non-adiabatic calculation of \cite{Klaiber2015}, 
the complex transition point $t_s$ found has a negative real component. 
Interestingly, the corresponding ionization time $t_i$, which is the first time of the trajectory on the real axis, is larger than zero, see figure 2d of \cite{Klaiber2015}. 
In consequence, this particular formalism predicts nonzero real time to pass while the photoelectron tunnels through the potential barrier.

A starting time $t_0$ before and corresponding ionization time $t_i$ after the peak of the laser field leads to a very intuitive picture of an optimization problem.
Assuming a photoelectron spends some finite time $\tau$ in the classically forbidden region, then the probability of the tunnelling process would be maximized if the integrated barrier width during $\tau$ is minimized.

In the attoclock method, a numerical value for $t_0$ was necessary so that the reference calculations, which assume zero tunnelling time, could be launched at the appropriate initial time. 
The estimate for $t_0$ is based on the instantaneous tunnelling assumption, and the fact that the tunnelling probability rate depends exponentially on the field strength, thus reacting very sensitively to even the slight changes in the field magnitude at large ellipticity.
Consequently, ionization would happen preferentially in the moments of maximal field strength, along the major axis of polarization.
Therefore, the SCT simulations were launched at
\begin{equation}
t_i = t_0 + \tau = t_0 + 0, 
\end{equation}
where $t_0$ was assumed to be the peak of the field. For a wave form as defined in \eqref{eq:LaserField} with $\phi_{\mathrm{CEO}} = 0$, this meant $t_0 = 0$. 
Therefore, the tunnelling times $\tau_A$ extracted from the attoclock experiment are in reference to this starting time $t_0 = 0$. 

However, based on the earlier discussion in this section, the physical $t_0$ possibly should be chosen before the peak of the field. 
Additionally, the instantaneous ionization rate analysis as presented in \cite{Ivanov2018} seems to exclude an asymmetric distribution of the tunnelling time such as $\tau = t_i - 0$ with respect to the laser field. 
None of the investigations mentioned above predict a numerical value for what $t_0$ should be in the particular case of a helium target in an elliptical laser field, in three dimensions. 
Therefore, we can not perform a quantitative analysis of the experimental data with a modified $t_0$ assumption. 
Nevertheless, we can state that any correction of $t_0$ from the peak to before the peak would lead to a larger extracted tunnelling times $\tau>\tau_A$ compared to the attoclock delay $\tau_A$ which is presented in figure \ref{fig:MyFieldDelayNA} for example.

\section{Summarized Influence on Attoclock Interpretation} \label{sec:Summary}

Looking at all these individual aspects of strong-field tunnel ionization, we can conclude the following.
\begin{figure}[ht]
	\centering
	\includegraphics*[width=0.8\textwidth]{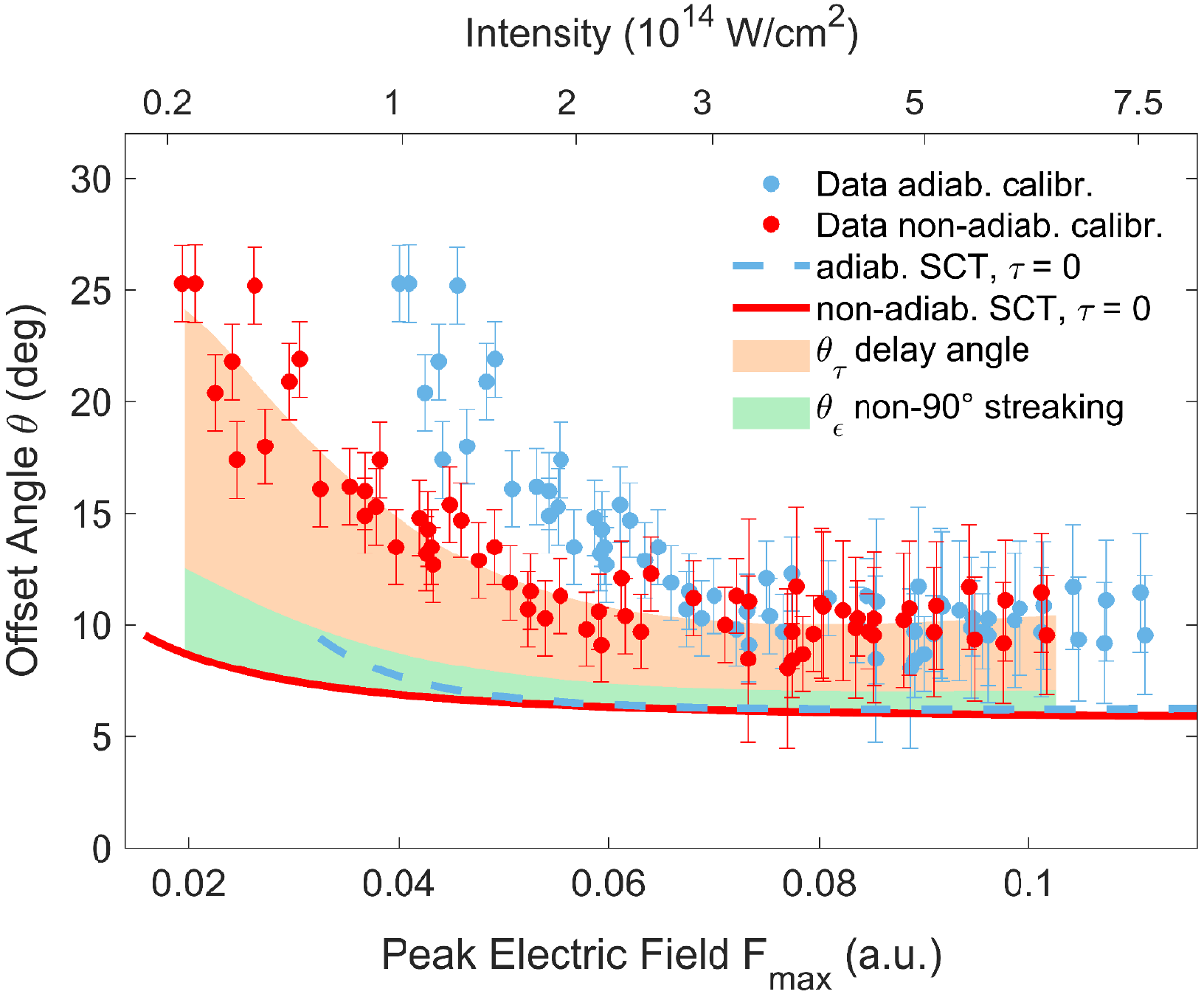} 
	\caption{Streaking offset angles of the recalibrated data set (red dots) compared to the original adiabatic field strength calibration data (blue dots). Also if non-adiabatic field strength calibration is used, an angle difference between the measured streaking angle $\theta$ and the zero tunnelling time prediction $\theta_{\mathrm{SCT}}$ calculated from single classical trajectories (SCT) remains. The offset angle difference $\theta - \theta_{\mathrm{SCT}}$ can be explained by a tunnelling delay time corresponding to $\theta_{\tau}$.}
	\label{fig:MyFieldAngle}
\end{figure}
Within the attoclock framework, the offset angle difference
\begin{equation}
\theta-\theta_{\mathrm{SCT}} = \theta_{\tau} + \theta_{\epsilon} \label{eq:theta_sums}
\end{equation}
(compare again figure \ref{fig:VMIdataSCT})
is explained as a tunnelling delay time (orange band in figure \ref{fig:MyFieldAngle}) $\theta_{{\tau}}$, plus an additional streaking angle $\theta_{\epsilon}$ (green band in figure \ref{fig:MyFieldAngle}). 
The $\theta_{\epsilon}$ is due to the elliptical polarization of the ionizing laser field.
Only when the electric field happens to point along either the major or minor axis of the polarization ellipse are the electric field vector and the vector potential orthogonal to each other.
So if a photoelectron enters the continuum at any time $t_i$ other than those precise moments, the ellipticity of the laser field leads to non-90 degree streaking angles, even in a purely field-driven case ignoring any other influences on the trajectory.
$\theta_{\epsilon}$ therefore depends on the ionization time $t_i$ at which a trajectory enters the continuum and the ellipticity of the driving field $\epsilon = 0.87$, and can be estimated as follows.

The total field-induced streaking angle
\begin{equation}
\theta_{\mathrm{field}} = \frac{\pi}{2} + \theta_{\epsilon},
\end{equation}
with the ellipticity correction $\theta_{\epsilon}$ to the 90 deg streaking angle is given by the angle between $\mathbf{F}(t_i)$ and  $\mathbf{A}(t_i)$.
Therefore, we can write
\begin{equation}
\cos(\theta_{\mathrm{field}}) = \frac{\mathbf{F}(t_i)\cdot\mathbf{A}(t_i)}{|F(t_i)||A(t_i)|} 
	= \frac{-\sin(\omega t_i) \cos(\omega t_i) + \epsilon^2 \cos(\omega t_i) \sin(\omega t_i)}{\sqrt{\cos^2(\omega t_i) + \epsilon^2 \sin^2(\omega t_i)}\sqrt{\sin^2(\omega t_i) + \epsilon^2 \cos^2(\omega t_i)}}
\end{equation}
Taking the Taylor expansions up to first order on both sides individually, for $\theta_{\mathrm{field}} \approx \frac{\pi}{2}$ and $t_i \approx 0$ respectively leads to
\begin{equation}
\theta_{\mathrm{field}} - \frac{\pi}{2} = \theta_{\epsilon} = \frac{(1-\epsilon^2)\omega t_i}{\epsilon}. \label{eq:theta_epsilon}
\end{equation}
The remaining angle difference $\theta_{{\tau}}$ is then interpreted as the time interval $\tau_A = t_i-t_0 = t_i - 0$ after the peak of the electric field until the electron exits the tunnelling barrier and enters the continuum 
\begin{equation}
\theta_{\tau} = \arctan\left( \frac{\epsilon \sin(\omega t_i)}{\cos(\omega t_i)}\right) \approx \epsilon \omega t_i \label{eq:theta_tau}
\end{equation}
Combining equations \eqref{eq:theta_sums}, \eqref{eq:theta_epsilon} and \eqref{eq:theta_tau}, the attoclock delay can finally be extracted as
\begin{align}
\theta-\theta_{\mathrm{SCT}} &= \frac{(1-\epsilon^2)\omega t_i}{\epsilon} + \epsilon \omega t_i \nonumber \\
\tau_A := t_i -0 &= \frac{\theta - \theta_{\mathrm{SCT}}}{\omega \left( \frac{1-\epsilon^2}{\epsilon} + \epsilon \right)}
\end{align}
from a measured streaking offset angle $\theta$ and a calculated zero-time reference $\theta_{\mathrm{SCT}}$.
Multi-electron effects do not significantly influence the final photoelectron momentum spectrum \cite{Emmanouilidou2015,Majety2017}, so the single active electron approximation for the single classical trajectory reference, obtaining $\theta_{SCT}$, is valid. 

\begin{figure}[ht]
	\centering
	\includegraphics[width=0.8\textwidth]{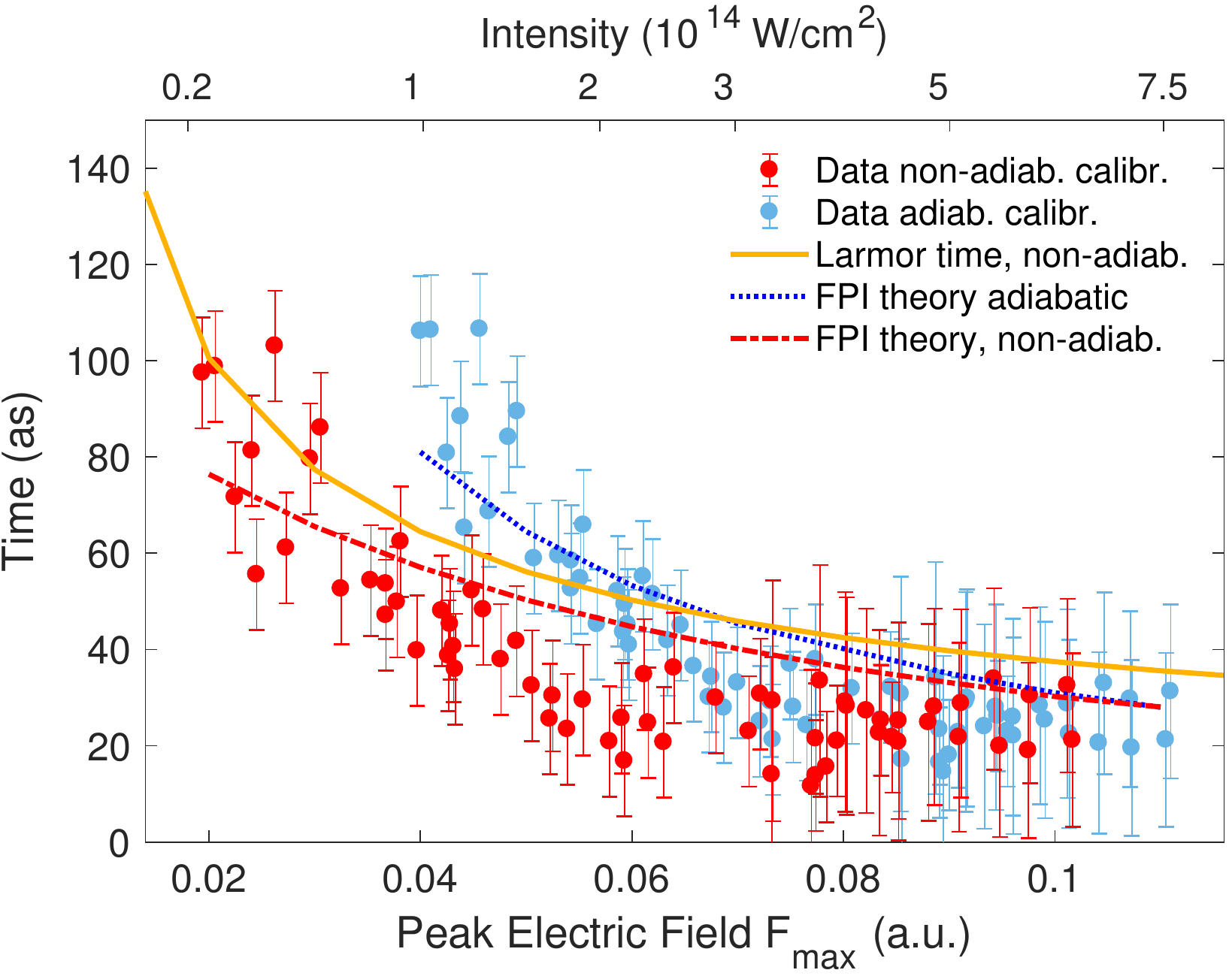}
	\caption{Extracted attoclock delay times $\tau_A$ corresponding to the non-adiabatically calibrated data (red dots), compared to adiabatically calibrated previous results (blue dots). The lines show the predictions of a Feynman Path Integral (FPI) calculation \cite{Landsman2014b} for both adiabatic and non-adiabatic barrier width (blue dotted and red dot-dashed respectively), as well as the Larmor time \cite{Buttiker1982,Buttiker1983,Landsman2015} (solid orange).}
	\label{fig:MyFieldDelayNA}
\end{figure}

Contrary to prior work \cite{Boge2013}, the SCT prediction in the fully non-adiabatic framework shows the same qualitative behaviour as in the adiabatic approximation, if all initial conditions of the classical trajectories are calculated non-adiabatically, see figures \ref{fig:My_Field_Angle_Ivanov2014} and \ref{fig:MyFieldAngle}. Consequently, the values of the extracted tunnelling delay times as defined in the attoclock method are comparable to the results published in \cite{Landsman2014,Landsman2014b}. 
However, these values are shifted to lower field strengths due to the calibration method including the initial transverse momentum predicted for elliptical polarization in the non-adiabatic case \cite{PPT2}, see figure \ref{fig:MyFieldDelayNA}.

Also theoretical predictions, or rather their evaluation, are affected by  non-adiabatic effects.  
The effective barrier width is comparatively shorter in the non-adiabatic framework. 
This has a noticeable influence on the Feynman Path Integral predictions for tunnelling time \cite{Landsman2015}, where the transmission wave function is evaluated at the calculated exit point. 
Both the adiabatic version as published in \cite{Landsman2014b} (blue dotted) as well as a non-adiabatic version (red dot-dashed) are plotted in figure \ref{fig:MyFieldDelayNA}.
The only difference between the two versions is the different exit radius, all other parameters of the calculation are identical.
The Larmor time is defined as \cite{Baz1967,Rybachenko1967}
\begin{equation}
\tau_{LM} = \frac{\partial \phi}{\partial V},
\end{equation}
where $\phi$ is the phase of the transmission amplitude through the potential barrier, and $V$ is the barrier height. 
Interestingly, the same non-adiabatic effect of a shorter exit radius only leads to a tiny shift, much smaller than the error bars of the data, for the Larmor time values. 
Therefore, figure \ref{fig:MyFieldDelayNA} only shows the values for the non-adiabatic case (orange solid line). 

\begin{figure}[ht]
	\centering
	\includegraphics[width=\linewidth]{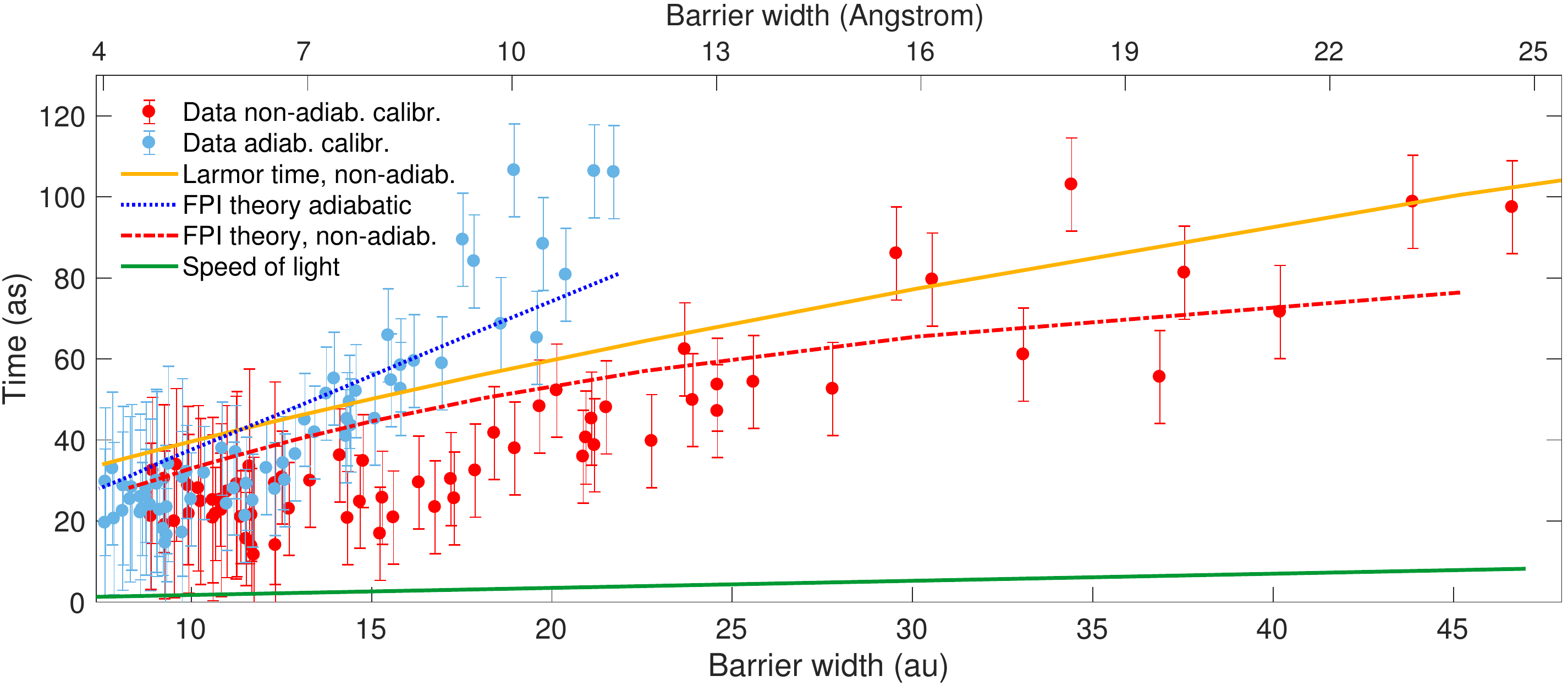}
	\caption{Extracted attoclock tunnelling delay times $\tau_A$  in the adiabatic (blue \cite{Landsman2014b}) and non-adiabatic version (red), compared to the corresponding Feynman Path Integral (FPI) estimates (blue dotted for the adiabatic and red dot-dashed for the non-adiabatic case) and the non-adiabatic version of the Larmor time (orange solid line). The speed-of-light (green solid line) is much faster than the extracted motion.}
	\label{fig:My_TimeBarrier}
\end{figure}

Of course, the extracted tunnelling times can also be plotted versus the length of the tunnelling barrier. 
For figure \ref{fig:My_TimeBarrier}, the barrier width $W$ was always estimated by the corresponding short-range potential width
\begin{equation}
W \approx \frac{\IP}{F_{\mathrm{max}}}.
\end{equation}
Since the non-adiabatic field strength calibration yields smaller values for the maximal field strengths for the same data sets, those corresponding barrier widths are significantly larger, meaning that the photoelectron travels a much larger distance in the same time as was originally deduced.
But still, the green solid line in figure \ref{fig:My_TimeBarrier} shows the values corresponding to a motion at speed-of-light. 
All the experimental data are significantly larger times than that, implying sub-luminal speed of the photoelectrons inside the potential barrier.

Looking at the longitudinal momentum distribution of the photoelectron wave packet at the tunnel exit, there are some results indicating that it should be a spread \cite{Bondar2008,Wavepacket,Hofmann2013,Hofmann2014} (compatible with the uncertainty principle) and might have a non-zero most probable value, pointing away from the ion \cite{Teeny2016,Camus2017}.
Proof-of-principle CTMC calculations however showed that any combination of these effects either only lead to insignificant shifts of the final angular PMD, or these shifts are essentially explained by the simple conservation of canonical momentum (figure \ref{fig:DeltaCoM}). 
On the other hand, the angular shift introduced by a non-zero most probable initial longitudinal momentum would reduce $\theta_{SCT}$, thereby increasing estimated tunnelling time.
Consequently, the streaking angle offset difference between experiment and $\theta_{\mathrm{SCT}}$ either stays the same or would only increase, leading to an even larger extracted tunnelling time $\tau_A$.
Last but not least, several publications either directly found a starting time before the peak of the electric field \cite{Klaiber2015,Teeny2016}, or their results suggest that this might be an option \cite{Ni2016,Torlina2015}.
This of course is another effect that acts to increase the extracted tunnelling time in experiments based on the attoclock idea \cite{Landsman2014b,Camus2017,Sainadh2017}. 
However, none of these approaches immediately yield a quantitative prediction of either the most probable initial longitudinal momentum, nor the starting time, for the case of a three-dimensional helium atom. 

\section{Conclusion and Outlook} \label{sec:Conclusion}

To summarize, a number of recent findings affect either the underlying semiclassical model or the data calibration in the attoclock experiment, such that this updated version finds different values for the tunnelling time than were originally published in \cite{Landsman2014b}. 
In particular, there is a shift of attoclock measured tunnelling delays to lower intensity values due to a shift in the experimental calibration of intensity when non-adiabatic transverse velocity at the tunnel exit is taken into account.
Many other approximations however were confirmed to be valid once again, such as the single active electron approximation, neglecting multi-electron effects. 
However, we were unable to find any effect or model that would render the experimental tunnelling time significantly smaller or even close to zero for the case under consideration.
Two more independent experiments have since been peer-reviewed and published, both also finding finite and real tunnelling time.
The analytical models used to explain these experiments are fully quantum (based on the Wigner approach) in the case of \cite{Camus2017}, and quasi-classical in the case of \cite{Fortun2016}. 

On the other hand, a vast range of theoretical approaches exists but often uses a different set of approximations, and even more crucially, different definitions of tunnelling time. 
Consequently, there is still no clear theoretical consensus.

Both the initial longitudinal momentum and the starting time before the peak can not be quantified yet for the helium target (or any larger atoms, for that matter). 
So we have to leave it at a qualitative statement for now, assuring that taking account of these effects should increase the extracted tunnelling times. 
This points to a need for further theoretical investigation in the quantum description of the strong field tunnel ionization process. 

Finally, it is important to recognize that exact definitions matter, and influence both the outcome and interpretation of a study. 
Most of the presented works use their own individual observables and definitions of the system under investigation. 
This makes direct comparisons a challenging task.
Nevertheless, experimental data can be quantitatively explained by models including some form of a finite tunnelling time, while most models assuming instantaneous tunnelling so far were not able to reproduce the measurements.  

For more definitive tests, it is desirable to do more comprehensive studies of atomic hydrogen, where multi-electron effects can be neglected.  While calculations for atomic hydrogen are more definitive, experimental measurements are considerably more challenging than the corresponding measurements on noble gases.  Longer wavelengths, which approach the adiabatic tunnelling regime, would also provide a more convincing test and allow for comparison with more non-adiabatic experiments.  On the analytic front, it is important to further explore the time-zero assumption as starting at the peak of the laser field.  Any change to this time-zero calibration would obviously have a direct impact on the extraction of tunnelling time from attoclock experiments.  

\section*{Funding}
A.S.L. acknowledges the Max Planck Centre of Attosecond Science (MPC-AS). This research was supported by the National Centre of Competence in Research Molecular Ultrafast Science and Technology (NCCR MUST), funded by the Swiss National Science Foundation (SNSF) and an ERC Advanced Grant (Attoclock-320401) within the seventh framework programme of the European Union.

\section*{Disclosure statement}
The authors declare no competing financial interests.


\clearpage
\bibliographystyle{tfp}
\bibliography{References}

\end{document}